\documentclass[aps,prd,twocolumn,amsfonts,amssymb,amsmath]{revtex4}

\usepackage{graphicx}

\newcommand{\bmath}[1]{\mbox{{\boldmath{{$#1$}}}}}

\begin{document}

\title{Higher-dimensional solitons and black holes with
a non-minimally coupled scalar field}

\author{Dominic Hosler}

\affiliation{
Department of Physics and Astronomy,
The University of Sheffield,
Hicks Building,
Hounsfield Road,
Sheffield.\
S3 7RH
United Kingdom
}

\author{Elizabeth Winstanley}

\email{E.Winstanley@sheffield.ac.uk}

\affiliation{
School of Mathematics and Statistics,
The University of Sheffield,
Hicks Building,
Hounsfield Road,
Sheffield.\
S3 7RH
United Kingdom
}

\begin{abstract}
We study higher-dimensional soliton and hairy black hole solutions
of the Einstein equations non-minimally coupled to a scalar field.
The scalar field has no self-interaction potential but a
cosmological constant is included.
Non-trivial solutions exist only when the cosmological constant is
negative and the constant governing the coupling of the scalar
field to the Ricci scalar curvature is positive.
At least some of these solutions are stable
when this coupling constant is not too
large.
\end{abstract}

%\date{\today }


\maketitle

\section{Introduction}
\label{sec:intro}

The existence and uniqueness of black hole solutions of the Einstein
equations with various types of matter has been a rich avenue of
research for many years.
The classic ``no-hair'' theorems (see, for example, \cite{heuslerreview} for
a review) proved the uniqueness of the Kerr-Newman family of metrics describing
four-dimensional, asymptotically flat, black hole solutions of the Einstein equations
with an electromagnetic field or in a vacuum.
More recently, there has been an explosion of interest in the generalization of these
uniqueness results to higher-dimensional black holes, and particular, in the
non-uniqueness of higher-dimensional rotating black holes
(see, for example, \cite{emparanreall}).

Black hole solutions of the Einstein-scalar field system have been
studied for almost as long as those of the Einstein-Maxwell system
(see \cite{heuslerreview} for a detailed review, and \cite{bekreview}
for a summary).
The case of a minimally coupled scalar field has been the most extensively
studied, with a number of ``no-hair'' results proved, in asymptotically
flat space, particularly for static, spherically symmetric black holes
\cite{mincoupleduniqueness}.
These results depend on some assumptions about the form of the self-interaction
potential, typically that it is positive semi-definite.
When these assumptions are not satisfied, asymptotically flat black hole and
soliton solutions of the field equations can be constructed,
some numerically with a particular choice of self-interaction potential
\cite{mincoupnum}, and there are also analytic solutions (often with unusual
potentials) \cite{mincoupan}, which can be generated from vacuum solutions
in the static case \cite{vacgen}.

Including a cosmological constant changes the picture for minimally
coupled scalar fields, providing the self-interaction potential is non-zero
\cite{toriidS,toriiadS}.
When the cosmological constant is positive, numerical \cite{toriidS} and analytic
\cite{zlosh} black hole solutions have been found
when the self-interaction potential is non-convex \cite{lahiri},
although at least some of these solutions are unstable \cite{toriidS}.
On the other hand, with a negative cosmological constant,
stable black holes with minimally coupled scalar field hair have been found numerically
\cite{toriiadS}, and there are also some analytic solutions
\cite{mtzminimal,farakos,zeng}.
Such asymptotically anti-de Sitter (adS) solutions have attracted much recent
interest in the literature due to their interpretation via the adS/CFT (conformal
field theory) correspondence (some works on this topic include
\cite{zeng,adSCFThair}), particularly for black holes with scalar hair in supergravity
theories \cite{SUGRAhair}, although the latter are unstable \cite{SUGRAhairstab}.

When non-minimal coupling of the scalar field to the Ricci scalar curvature is
included in the model, the case of conformal coupling has been of particular interest.
The BBMB solution \cite{BBMB} is the unique \cite{BBMBunique}
static, asymptotically flat, black hole solution of the field
equations in four space-time dimensions,
when the self-interaction potential is zero, but is unstable
\cite{BBMBunstable}.
Furthermore, the scalar field diverges on the event horizon.
An analytic solution has been found in asymptotically flat space when the
scalar field self-interaction potential is non-zero \cite{conformalnonzeroV}.
The analogue of the BBMB solution in four dimensions
when there is a positive cosmological constant
has a quartic self interaction potential \cite{mtz}.
Although the scalar field does not suffer from the
divergences of the BBMB black hole, it is still unstable \cite{harper}.
As for minimal coupling, in the presence of a negative cosmological constant,
four-dimensional
stable solitons \cite{raduew} and black holes \cite{ew2003} have been found
when the self-interaction potential is either zero or quadratic.

The case of non-minimal coupling which is not conformal coupling has received less
attention in the literature.
In four-dimensional, asymptotically flat, space-time, various
black hole ``no-hair'' theorems
have been proved \cite{nonminnohair,saa}.
With a zero self-interaction potential, four-dimensional black holes are studied in
\cite{ew2005}.
It is found that non-minimally- and non-conformally-coupled scalar field hair
can be supported by the black hole only when there is a negative cosmological constant,
in agreement with numerical work \cite{pena}.
Furthermore, the hair is stable only when the constant governing the coupling between
the scalar field and the Ricci scalar curvature is such that the Breitenlohner-Freedman
bound \cite{BFbound} on the ``effective mass'' is satisfied.

Most of the above work is concerned with static, spherically symmetric black holes
in four space-time dimensions.
In asymptotically adS space, as well as spherically symmetric black holes,
the well-known topological black holes exist (see, for example, \cite{topological}).
Topological black holes with scalar field hair have not been widely studied,
apart from an analytic solution in the minimally coupled case \cite{4dtopnegcurvmincoup},
and numerical solutions in the conformally coupled case \cite{raduew}.

Solitons and black holes with scalar hair in more than four space-time dimensions
have received more attention in the literature in recent years, particularly
in asymptotically adS space \cite{farakos,SUGRAhair,raduew}.
The techniques for generating minimally coupled scalar solutions from solutions
of the vacuum Einstein equations have been generalized \cite{higherdimgen},
yielding new analytic solutions.
For conformally coupled scalar fields,
there is no higher-dimensional analogue of the asymptotically flat
BBMB black hole \cite{klimcik},
although the corresponding analytic solution in three-dimensional,
asymptotically adS space does exist \cite{3Dconf},
and there are numerical soliton and black hole solutions in adS in higher dimensions
\cite{raduew}, all with vanishing self-interaction potential.
A set of analytic solutions in various dimensions with a non-zero self-interaction
potential and a cosmological constant has also been considered recently
in the conformally coupled case \cite{nadalini}.
For other non-minimal coupling, Saa \cite{saa} has proved a no-hair theorem
in the asymptotically flat case, although for certain values of the
coupling constant assumptions on the magnitude of the scalar field are required.

Our purpose in this paper is to consider non-minimal, non-conformal coupling,
extending the analysis of \cite{ew2005} to solitons,
higher-dimensional black holes, and topological black holes in asymptotically
adS space. The structure of this paper is as follows.
In Sec.~\ref{sec:model} we describe our scalar field model, the field
equations and boundary conditions to be satisfied by soliton and black hole solutions,
and the conformal map which transforms our non-minimally coupled scalar field
system to one with a minimally coupled scalar field and a non-zero self-interaction
potential.
For most of the parameter space, we are able to prove that non-trivial solutions
cannot exist, and these non-existence results are presented in
Sec.~\ref{sec:no-hair}.
Non-trivial solutions exist in the remainder of the parameter space, and in
Sec.~\ref{sec:nontrivial} we present our numerical solutions and study
their thermodynamics and stability.
Finally our conclusions are presented in Sec.~\ref{sec:conc}.

\bigskip

\section{The model}
\label{sec:model}

\subsection{Field equations}
\label{sec:fieldeq}

We consider the following action, describing a
scalar field $\phi$ with non-minimal coupling to gravity in $n$ space-time dimensions:
\begin{equation}
S=
\frac {1}{2}
\int d^{n}x \, {\sqrt {-g}} \left[
\left( R -2\Lambda \right)
- \left( \nabla \phi \right) ^{2}
- \xi R \phi ^{2} \right] ,
\label{eq:action}
\end{equation}
where $R$ is the Ricci scalar curvature, and we have included a cosmological constant
$\Lambda $.
Here and throughout this paper, the metric has signature $(-,+,+,+)$ and we use units
in which $c=\hbar=8\pi G=k_{B}=1$.
The constant $\xi $ governs the coupling between the scalar field and the Ricci
scalar curvature.
Two values of $\xi $ are particularly important: minimal coupling $\xi = 0$,
and conformal coupling $\xi = \xi_c = (n-2)/[4(n-1)]$.
In the action (\ref{eq:action}) we have set any self-interaction potential to zero,
for simplicity.

Varying the action (\ref{eq:action})
with respect to the field variables,
we obtain the Einstein equations
\begin{widetext}
\begin{equation}
\left( 1 - \xi \phi ^2 \right) G_{\mu \nu} + g_{\mu \nu} \Lambda
 =  \left( 1 - 2 \xi \right) \nabla_{\mu} \phi \nabla_{\nu} \phi
 + \left( 2 \xi - \frac{1}{2} \right) g_{\mu \nu} \left(\nabla \phi \right)^2
 - 2 \xi \phi \nabla_{\mu} \nabla_{\nu} \phi
 + 2 \xi g_{\mu \nu} \phi \nabla ^2 \phi ;
\label{eq:einst}
\end{equation}
\end{widetext}
and scalar field equation
\begin{equation}
\nabla ^2 \phi - \xi R \phi = 0 .
\label{eq:scalarfieldeq}
\end{equation}
Taking the trace of the Einstein equations (\ref{eq:einst}) gives the Ricci scalar
curvature
\begin{equation}
R = - \frac
{2 \left( n - 1 \right) \left( \xi - \xi_c \right) \left( \nabla \phi \right) ^2
%+ 2 \xi \left( n - 1 \right) \phi \frac{\mbox{d}V}{\mbox{d}\phi}
- n \Lambda }
{\frac{n}{2} - 1 + 2 \left( n - 1 \right) \xi \left( \xi - \xi_c \right) \phi^2} .
\label{eq:ricciscalar}
\end{equation}

We are interested in static, spherically symmetric, solitons and black holes
for all values of the cosmological constant $\Lambda $;
and static topological black holes when $\Lambda <0$.
We therefore use the following ansatz for the metric,
in the usual Schwarzschild-like co-ordinates:
\begin{equation}
ds^2 =
- H(r) e^{2 \delta(r)} \, dt^2 +
H(r)^{-1} \, dr^2  + r^2 \, d \sigma ^2 _{n-2,k} ,
\label{eq:ansatz}
\end{equation}
where
\begin{equation}
d\sigma^2_{n-2,k} = d\psi^2 + f^2_k(\psi)\, d\Omega^2_{n-3}
\label{eq:dsigma}
\end{equation}
denotes the line element of an $(n-2)$-dimensional space $\Sigma_k$
with constant curvature.
The discrete parameter $k$ takes the values $1, 0, -1$, for which
we have the following forms of the function $f_k(\psi)$:
\begin{equation}
f_k(\psi) = \left\{
\begin{array}{ll}
\sin \psi, &\, \mbox{for } k=1,\\
\psi, &\, \mbox{for } k = 0,\\
\sinh \psi, &\, \mbox{for } k =-1.
\end{array} \right.
\label{eq:kform}
\end{equation}
The value $k=1$ gives the usual spherically symmetric metric, with the
hypersurface $\Sigma_1$ equal to an $(n-2)$-sphere; when $k=-1$, the hypersurface
$\Sigma _{-1}$ has constant negative curvature, and when $k=0$, the hypersurface
$\Sigma _{0}$ is an $(n-2)$-dimensional Euclidean space (see \cite{topological}
for further details).
It is convenient to introduce a new metric function $m(r)$ by
\begin{equation}
H(r) = k - \frac{2 m(r)}{r^{n-3}} - \frac{2\Lambda r^{2}}{(n-2)(n-1)}.
\label{eq:metricfunc}
\end{equation}

\begin{widetext}
With the ansatz (\ref{eq:ansatz}), and assuming that the scalar field $\phi $
depends only on the radial co-ordinate $r$,
the field equations (\ref{eq:einst}, \ref{eq:scalarfieldeq}) take the form:
\begin{eqnarray}
0 & =  & \frac{n-2}{2r}(1-\xi \phi^2)\left[H'
-\frac{n-3}{r}(k-H)\right] -
\left(2\xi-\frac{1}{2}\right)H\phi'^2+\xi \phi \phi'(H'+2H
\delta') -2\xi \phi\nabla^2 \phi +\Lambda ;
\nonumber \\
0 & = &
\frac{n-2}{r}(1-\xi\phi^2)\delta'-(1-2\xi)\phi'^2-2\xi\phi(\delta'\phi'-\phi'') ;
\nonumber
\\
0 & = &
H \phi''+\phi'\left(H\delta'+H'+H\frac{n-2}{r}\right)
-\xi R \phi.
\label{eq:fieldeqs}
\end{eqnarray}

We now summarize the discussion of Sec.~\ref{sec:intro},
namely what is known about the existence of soliton and black hole
solutions of these field equations, for various values of the cosmological constant
$\Lambda $ and coupling constant $\xi $.
Firstly, we bring together in one table the known results for solitons and black holes
in four space-time dimensions:
\begin{table}[h]
%\begin{ruledtabular}
\begin{tabular}[c]{|c||c|c|c|c|c|}
\hline
$n=4$ & $\xi < 0$ & $\xi = 0$ & $0 < \xi < \xi_c$ & $\xi = \xi_c$ & $\xi > \xi_c$ \\
\hline
\hline
$\Lambda > 0$ & no black
& no black
& no black
& no black
& no black \\
& hole hair \cite{ew2005}
& hole hair \cite{toriidS,lahiri}
& hole hair \cite{ew2005}
& hole hair \cite{ew2003}
& hole hair \cite{ew2005}
\\
\hline
$\Lambda = 0$
& no black
& no black
& no black
& unstable black
& no black
\\
& hole hair \cite{nonminnohair}
& hole hair \cite{mincoupleduniqueness}
& hole hair \cite{nonminnohair}
& hole hair \cite{BBMB}
& hole hair \cite{nonminnohair}
\\
\hline
$\Lambda < 0$
& no black
& no black
& stable black
& stable solitons
& unstable black  \\
%\hline
& hole hair \cite{ew2005}
& hole hair \cite{toriiadS}
& hole hair \cite{ew2005}
& and black holes \cite{raduew,ew2003}
& hole hair \cite{ew2005}
\\
\hline
\end{tabular}
%\end{ruledtabular}
\caption{Summary of existence and non-existence of soliton and black hole
solutions in four space-time dimensions.}
\label{tab:PreviousKnowledgeForV0N4}
\end{table}
\newline
It can be seen that the existence of hair
for spherically symmetric black holes has been studied for all values
of the cosmological constant $\Lambda $ and coupling constant $\xi $.
However, to date solitons and topological black holes have been considered only
when the scalar field is conformally coupled, $\xi = \xi _{c}$.

A similar table for $n>4$ reveals that the parameter space has been much less widely
explored:
\begin{table}[h]
%\begin{ruledtabular}
\begin{tabular}[c]{|c||c|c|c|c|c|}
\hline
$n>4$ & $\xi < 0$ & $\xi = 0$ & $0 < \xi < \xi_c$ & $\xi = \xi_c$ & $\xi > \xi_c$ \\
\hline
\hline
$\Lambda > 0$ &
&
&
&
&
\\
\hline
$\Lambda = 0$
& no black hole hair \cite{saa}
&
& no black hole hair \cite{saa}
& no black hole hair \cite{klimcik}
& no black hole hair \cite{saa}
\\
\hline
$\Lambda < 0$
&
&
&
& stable solitons and black holes \cite{raduew}
&
\\
\hline
\end{tabular}
%\end{ruledtabular}
\caption{Summary of existence and non-existence of soliton and black hole
solutions in more than four space-time dimensions.}
\label{tab:PreviousKnowledgeForV0}
\end{table}
\newline
It should be emphasized that the results in Tabs.~\ref{tab:PreviousKnowledgeForV0N4}
and \ref{tab:PreviousKnowledgeForV0}
are only for {\em {zero}} self-interaction potential, as many of the solutions
which have been found in more than four space-time dimensions have a non-zero potential
$V$ \cite{farakos,SUGRAhair,nadalini}.
The aim of the present paper is to complete
table \ref{tab:PreviousKnowledgeForV0}.
\end{widetext}

\subsection{Boundary conditions}
\label{sec:boundary}

Before we can investigate the existence (or otherwise) of black hole and soliton
solutions of the field equations (\ref{eq:einst}, \ref{eq:scalarfieldeq}) we need
to specify the boundary conditions at the origin (for solitons), black hole
event horizon (if there is one) and at infinity.

Firstly, regular soliton solutions can exist only when $k=1$.
Near the origin, the field variables have the expansions~\cite{raduew}:
\begin{eqnarray}
H & = & 1
+\frac{2 \left( \Lambda - 2 \xi^2 \phi_0^2 R_0 \right)}{(n-1)(n-2)(\xi \phi_0^2 -1)}
r^2 + O(r^{4});
\nonumber \\
\delta & = & \delta_0
+ \frac{\xi^2 \phi_0^2 R_0}{(n-1)(n-2)(\xi \phi_0^2 -1)} r^2 + O(r^4);
\nonumber \\
\phi & = & \phi_0 + \frac{\xi \phi_0 R_0}{2 (n-1)} r^2 + O(r^4);
\label{eq:rto0}
\end{eqnarray}
where
\begin{equation}
\label{eq:R0}
R_0 = \frac{n \Lambda}{\frac{n}{2} - 1 + 2 (n-1) \xi (\xi - \xi_c) \phi_0^2}
\end{equation}
is the value of the Ricci scalar curvature at the origin.
We note that the constants $\delta _{0}$ and $\phi _{0}$ are arbitrary; the value
of $\delta _{0}$ will be fixed by the boundary conditions at infinity.

For black holes with a regular, non-extremal event horizon at $r=r_{h}$, we assume
that all the variables have regular Taylor series expansions in a neighbourhood
of the event horizon:
\begin{eqnarray}
\nonumber
H(r)&=&H'(r_h)(r-r_h)+O\left(r-r_h\right)^2;
\\
\nonumber
 \delta(r) &  = & \delta_h+\delta'(r_h)(r-r_h)+O\left(r-r_h\right)^2;
\\
\phi(r)&=&\phi_h+\phi'(r_h)(r-r_h)+O\left(r-r_h\right)^2;
\label{eqar:rhfuncs}
\end{eqnarray}
where \cite{raduew}
\begin{eqnarray}
\nonumber
H'(r_h)&=&(n-3)\frac{k}{r_h}+\frac{2r_h\big(\Lambda
-\xi^{2}\phi_h^{2}R_h\big)}{(n-2)(\xi \phi_h^2-1)};
\\
\phi'(r_h) & = & \frac{\xi R_{h} \phi_h}{H'(r_h)};
\label{eqar:rhderivs}
\end{eqnarray}
and
\begin{equation}
\label{eq:Ratrh}
R_h=\frac{n\Lambda}{n/2-1+2(n-1)\xi(\xi-\xi_c)\phi_h^2}
\end{equation}
is the value of the Ricci scalar curvature on the event horizon.
We computed $\delta '(r_{h})$ directly (without using $\phi ''(r_{h})$
as in \cite{raduew}), but the expression so obtained is too lengthy to
reproduce here.
Since the field equations do not involve $\delta $, but only its derivatives,
the constant $\delta _{h}$ is arbitrary, and will be fixed by the boundary conditions
at infinity.
The constant $\phi _{h}$ is also arbitrary.
Similar expressions to (\ref{eqar:rhfuncs}, \ref{eqar:rhderivs}) hold at the cosmological
horizon in the case $\Lambda >0$.

We also require the behaviour of the field variables as $r\rightarrow \infty $.
To ensure
consistency between the Einstein equations (\ref{eq:fieldeqs})
and the form of the Ricci scalar (\ref{eq:ricciscalar}),
since the self-interaction potential is zero,
we will assume
that the scalar field vanishes as $r\rightarrow \infty $.

Considering firstly the asymptotically flat case $\Lambda = 0$,
the field variables have the following behaviour:
\begin{eqnarray}
\phi (r) & = &  \frac {c_{1}}{r^{n-3}}
+ O \left( r^{-n+2} \right) ;
\nonumber
\\
H(r) & = & 1 + O(r^{-n+3}) ;
\nonumber
\\
\delta (r) & = & O \left( r^{-2(n-3)} \right) ;
\label{eq:afinfinity}
\end{eqnarray}
and the Ricci scalar curvature is $R\sim O(r^{-2(n-2)})$.
The asymptotic form (\ref{eq:afinfinity}) means that the metric function $m(r)$
(\ref{eq:metricfunc}) tends to a constant as $r\rightarrow \infty $, namely the mass
of the solution.

For asymptotically anti-de Sitter solutions with $\Lambda <0$, the boundary
conditions near infinity for a massive, minimally coupled, scalar field have been
studied in depth \cite{henneaux}.
In our non-minimally coupled case, with $\Lambda \neq 0$,
we define a constant $p$ such that the behaviour of the scalar field as
$r\rightarrow \infty $ is
\begin{equation}
\phi(r) = \frac{c_1}{r^p} + O(r^{-(p+1)}),
\label{eq:phiinf}
\end{equation}
then from the scalar field equation (\ref{eq:fieldeqs}) we find
\begin{equation}
p = \left( \frac{n-1}{2} \right) \left( 1 \pm \sqrt{1 - \frac{4 n}{n-1} \xi} \right) .
\label{eq:p}
\end{equation}
The value of $p$ in (\ref{eq:p}) has the appropriate limits when
$\xi = \xi_{c}$ \cite{raduew} or $n=4$ \cite{ew2005}.
If $\xi <0$, then one of the values of $p$ is negative, leading to a scalar field
which diverges as $r\rightarrow \infty $.
We wish to rule this out and therefore choose the positive sign in (\ref{eq:p}) in
this case.
If $0\le \xi < (n-1)/(4n)$, both roots for $p$ are positive, and the dominant behaviour
of $\phi $ will be given by the smaller root, which has the negative
sign in (\ref{eq:p}).
If $\xi > (n-1)/(4n)$, the constant $p$ has a non-zero imaginary part, which leads
us to expect oscillatory behaviour in the function $\phi $ as $r\rightarrow \infty $,
as was observed in the four-dimensional case \cite{ew2005}.

The behaviour of the metric function $\delta (r)$ as $r\rightarrow \infty $ is readily
found, from the second equation in (\ref{eq:fieldeqs}), to be
\begin{equation}
\delta(r) = \frac{\delta_1}{r^{2p}} + O(r^{-2p-1}),
\label{eq:rinfdelta}
\end{equation}
while the leading order behaviour of the metric function $H(r)$ is
found, from the first equation in (\ref{eq:fieldeqs}):
\begin{equation}
H(r) = k - \frac {2\Lambda r^{2}}{(n-2)(n-1)} + O (r^{-2(p-1)}).
\label{eq:rinfH}
\end{equation}
The behaviour (\ref{eq:rinfdelta}) and (\ref{eq:rinfH}) ensures that the metric
(\ref{eq:ansatz}) approaches adS or de Sitter space as $r\rightarrow \infty $.

The behaviour of the metric function $H(r)$ (\ref{eq:rinfH}) has interesting
consequences for the metric function $m(r)$ (\ref{eq:metricfunc}).
We find
\begin{equation}
m(r) = M + \frac{M_1}{r^q} + O(r^{-q-1}),
\label{eq:mansatz}
\end{equation}
where
\begin{equation}
q = 2p + 1 - n = \pm (n-1) \sqrt{1 - \frac{4 n}{n-1} \xi} ,
\label{eq:q}
\end{equation}
and $M$ and $M_1$ are constants.
When $\xi <0$, the plus sign in (\ref{eq:q}) is relevant and $q$ is real and
positive, so that $m(r)$ converges as $r\rightarrow \infty $.
When $0<\xi < (n-1)/(4n)$, the negative sign in (\ref{eq:q}) will be the dominant
behaviour, so $q$ is real and $q<0$.
When $\xi > (n-1)/(4n)$, we have that $q$ is purely imaginary.
Thus the function $m(r)$ diverges as $r\rightarrow \infty $ for $0<\xi <(n-1)/(4n)$
and is oscillatory (with no decay in the oscillations) for $\xi > (n-1)/(4n)$.
Therefore, although the metric (\ref{eq:ansatz}) approaches pure adS or de Sitter as
$r\rightarrow \infty $, the function $m(r)$ has no finite limit.
We expect that the definition of the total mass of the
soliton or black hole, using the conserved charges approach of \cite{henneaux},
will have a contribution from the scalar field as well as from the metric function
$m(r)$.
We will not examine this further in this paper, leaving open the question of
how a finite mass for the configurations is defined.

However, it is not immediately apparent from (\ref{eq:mansatz}, \ref{eq:q})
how a finite limit for $m(r)$ arises in the case of conformal coupling \cite{raduew},
$\xi = \xi _{c} = (n-2)/[4(n-1)]$, when $q=-1$ (\ref{eq:q}).
To see this, we need to consider the constant $M_{1}$.
By comparing the form of the Ricci scalar expressed in terms of the scalar field
(\ref{eq:ricciscalar}) with that computed directly from the metric (\ref{eq:ansatz}),
we find the following equation for $M_{1}$:
\begin{eqnarray}
-q(q-1)\frac{3}{c_1^2\Lambda } M_1
\nonumber \\
& & \hspace{-2cm}
= n (n-1)^2 (\xi - \xi_c) \left[ \frac{2\xi}{n-2} - \frac{2p^2}{(n-1)(n-2)n} \right]
\nonumber \\
& & \hspace{-1.5cm}
- (2p-n)\frac{p}{n-1} \left[ p (1-4\xi) - 2\xi \right] .
\label{eq:M1sign}
\end{eqnarray}
From the expression for $p$ it is straightforward to show that the right-hand-side of
(\ref{eq:M1sign}) vanishes for all $n$ when $\xi = \xi_c$, while the coefficients
multiplying $M_{1}$ are non-zero in this case.
Therefore it must be the case that $M_{1}=0$ for $\xi =\xi _{c}$,
which means that the leading order term in $m(r)$ as $r\rightarrow \infty $
becomes the constant $M$, in agreement with \cite{raduew}.

\subsection{Conformal transformation}
\label{sec:conftrans}

As in four space-time dimensions \cite{ew2005}, we will find it useful
at various points in our analysis to employ a conformal transformation
\cite{BBMB,maeda} which maps the Einstein-non-minimally-coupled-scalar field system
to the mathematically simpler Einstein-minimally-coupled-scalar field system
(see \cite{otherconf} for related works on the conformal transformation).

\begin{figure}
\includegraphics[width=6cm,angle=270]{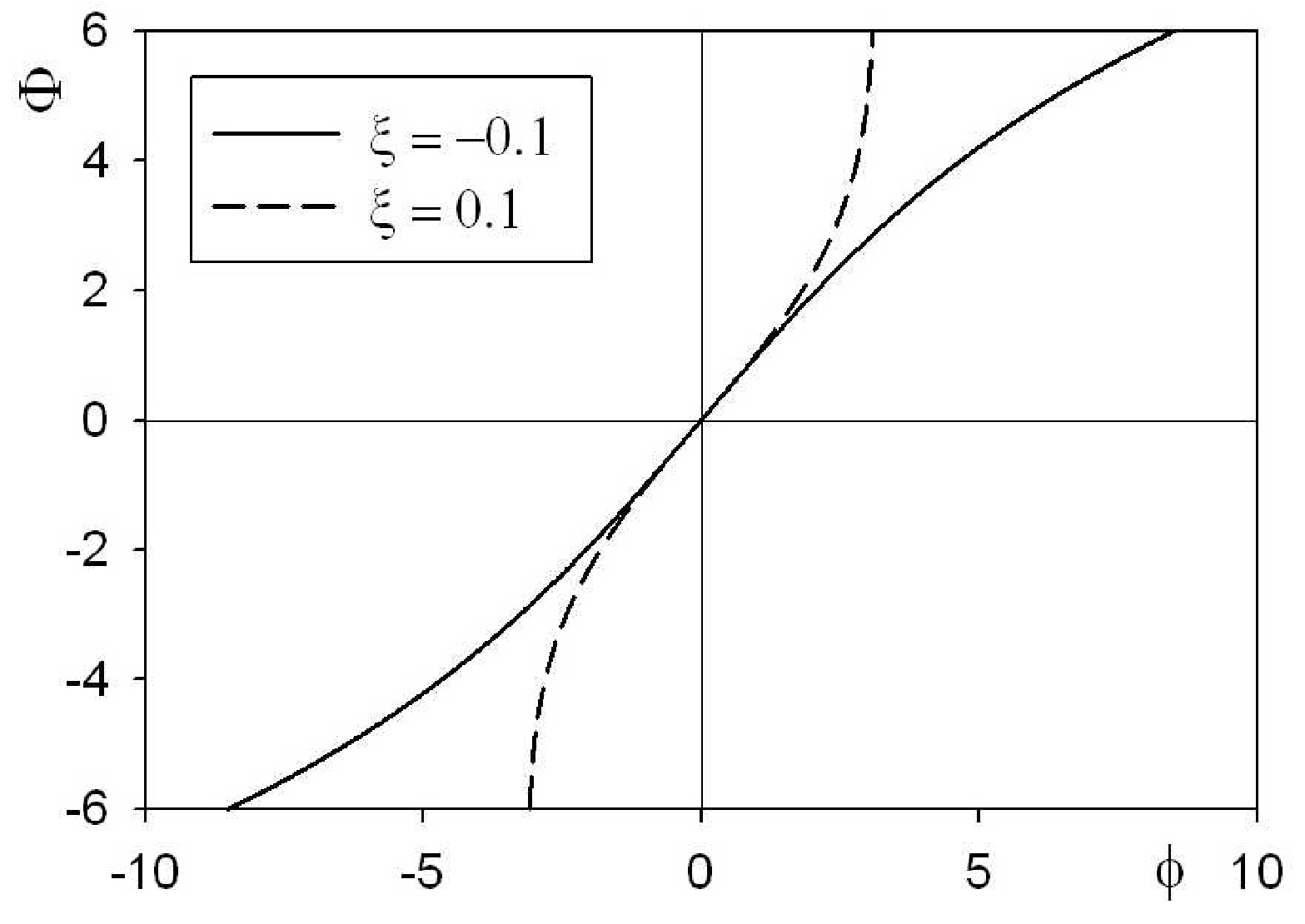}
\caption{The minimally coupled scalar field $\Phi $ (\ref{eq:conftransscal})
as a function of the non-minimally coupled scalar field $\phi $ for $n=6$
and two values of the coupling constant $\xi $.
Similar behaviour is observed for different values of $n$ and $\xi $.}
\label{fig:Phiphi}
\end{figure}

The conformal transformation is defined by the following \cite{maeda},
\begin{equation}
{\overline{g}}_{\mu\nu} = \Omega^{\frac{2}{n-2}} g_{\mu\nu},
\label{eq:conftransgmunu}
\end{equation}
where $\Omega = 1 - \xi \phi^2$.
This is only valid for $\Omega \ne 0$, which is always true for $\xi <0$
but bounds the scalar field when $\xi >0$.
With this transformation, the action (\ref{eq:action}) becomes
\begin{equation}
{\overline {S}} = \int d^{n}x \, {\sqrt {-{\overline{g}}}} \left[
\frac {1}{2}\left( {\overline {R}} -2\Lambda \right)
-\frac {1}{2} \left( {\overline {\nabla }} \Phi \right) ^{2}
- U(\Phi ) \right] ,
\label{eq:confaction}
\end{equation}
where a bar denotes quantities calculated using the transformed metric
(\ref {eq:conftransgmunu}), and the cosmological constant $\Lambda $ is
unchanged by the transformation.
As in \cite{maeda}, we define a new scalar field $\Phi $ by
\begin{equation}
\Phi = \int d\phi
\left[ \frac{(n-2)\Omega + (n-1)(4 \xi^2 \phi^2)}{(n-2) \Omega^2}
\right]^{ \frac{1}{2} } ,
\label{eq:conftransscal}
\end{equation}
choosing the constant of integration so that $\Phi =0$ when $\phi =0$.
The integral (\ref{eq:conftransscal}) can be performed analytically, however
the expression is very long except in the conformally coupled
case \cite{raduew}.
The new scalar field $\Phi $ (\ref{eq:conftransscal}) is a single-valued function
of $\phi $ as long as the quantity under the square root is positive,
which places a further constraint on the values of $\phi $.
The scalar field $\Phi $ is plotted as a function of $\phi $ in Fig.~\ref{fig:Phiphi}
for $n=6$ and two particular values of the coupling constant $\xi $.
When $\xi <0$, the function $\Phi $ is defined for all values of $\phi $, but when
$\xi >0$, the function $\Phi $ is defined only for those values of $\phi $ such that
the integrand in (\ref{eq:conftransscal}) is real, which gives a finite range of
values of $\phi $ centred on the origin.

\begin{figure}
\includegraphics[width=6cm,angle=270]{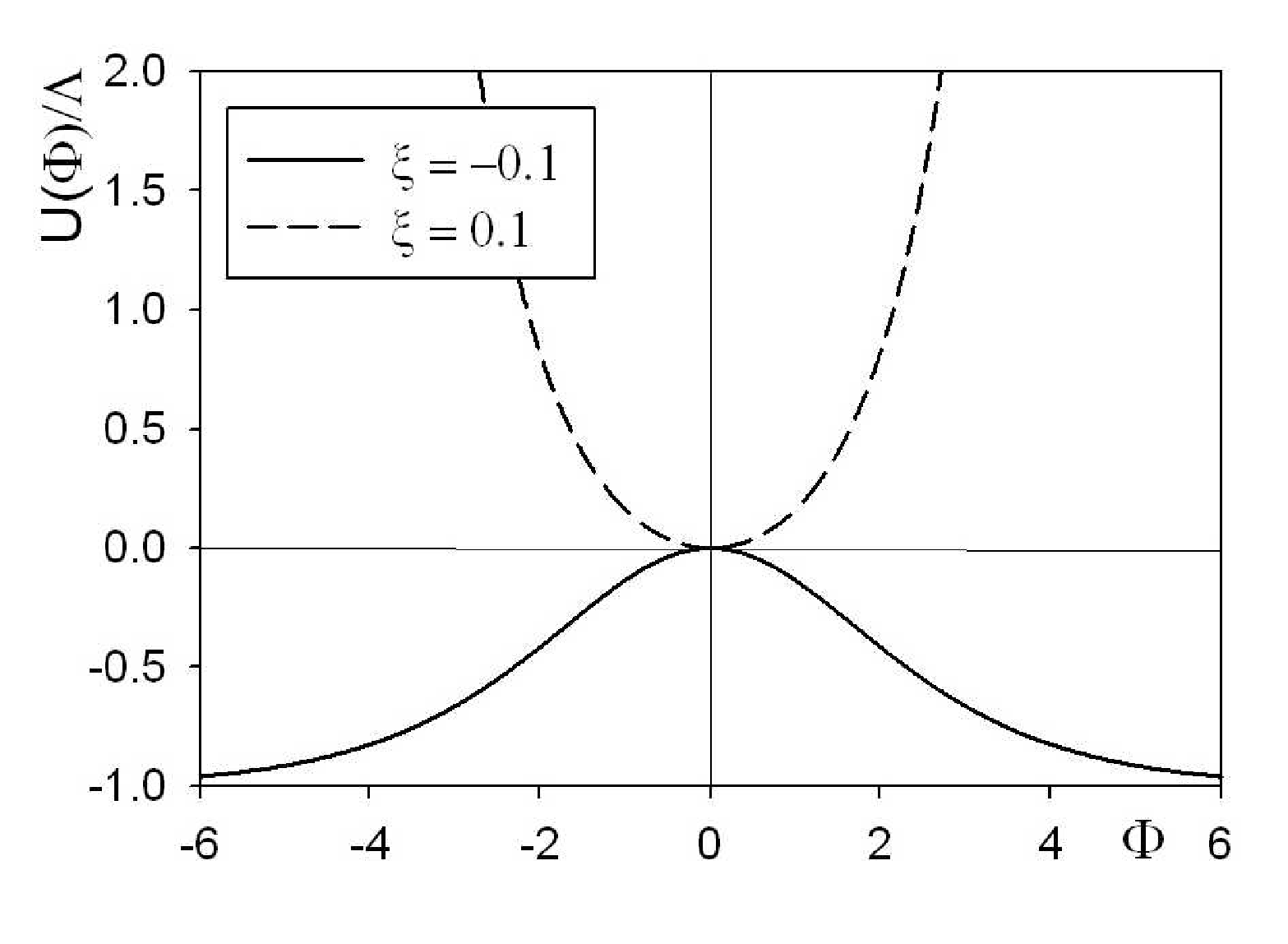}
\caption{The potential $U(\Phi )/\Lambda $ (\ref{eq:conftransfinalU})
as a function of the minimally coupled scalar field $\Phi $
for $n=6$ and two values of the coupling constant $\xi $.
Similar behaviour is observed for other values of $n$ and $\xi $.}
\label{fig:UPhi}
\end{figure}

The transformed potential $U(\Phi )$ can be written implicitly in terms of $\phi $
\cite{maeda}:
\begin{equation}
U = \Lambda \left[ \Omega^{-\frac{n}{n-2}} -1 \right] .
\label{eq:conftransfinalU}
\end{equation}
As in \cite{raduew,ew2005}, the presence of a cosmological constant means that
the potential $U(\Phi )$ for the minimally coupled scalar field $\Phi $ is non-zero
although the potential for the non-minimally coupled scalar field $\phi $ vanishes.
It should be emphasized that the potential $U(\Phi )$ is not physical,
which means that we can circumvent the conditions on the potential in the ``no-hair''
theorems in the minimally coupled case \cite{mincoupleduniqueness} and
have non-trivial solutions.
Examples of the shape of the potential $U(\Phi )$ can be seen in Fig.~\ref{fig:UPhi}.
For $\xi <0$, the potential is defined for all values of $\Phi $ and we find
that $U(\Phi )/\Lambda $ is negative for all $\Phi $, with a maximum at $\Phi =0$.
For $\xi >0$, the range of values of $\Phi $ is finite and we find that
$U(\Phi )/\Lambda $ is positive, with a minimum at $\Phi =0$.

The main advantage of the conformal transformation is that it simplifies the
field equations (\ref{eq:fieldeqs}) considerably.
If we assume that the transformed metric (\ref{eq:conftransgmunu}) is
spherically symmetric, we may write it in the form
\begin{equation}
d{\overline {s}}^2 =
- {\overline {H}}({\overline {r}}) e^{2 {\overline {\delta}}({\overline {r}})} \,
dt^2 +
{\overline {H}}({\overline {r}})^{-1} \, d{\overline {r}}^2
+ {\overline {r}}^2 \, d \sigma ^2 _{n-2,k} ,
\label{eq:confmetric}
\end{equation}
where we have introduced a new radial co-ordinate ${\overline {r}}$ by
\begin{equation}
{\overline{r}} = \Omega^{\frac{1}{n-2}} r ,
\end{equation}
and the new metric functions ${\overline {H}}({\overline {r}})$
and ${\overline {\delta }}({\overline {r}})$ are given by:
\begin{eqnarray}
{\overline{H}}({\overline{r}}) & = &
H(r) \left[1 - \frac{2\xi \phi \phi' r}{(n-2) \Omega} \right]^2 ;
\nonumber \\
{\overline{\delta}}({\overline{r}})
& = &
\delta(r) + \ln \left[ \Omega^{\frac{1}{n-2}}
\left(1 - \frac{2 \xi \phi \phi' r}{(n-2)\Omega} \right)^{-1}\right] .
\label{eqar:conftrans}
\end{eqnarray}
We note that ${\overline{H}} =0$ whenever $H=0$ so the horizon structure of the
space-time is preserved by the conformal transformation.
With the metric ansatz (\ref{eq:confmetric}), the field equations for the
transformed system take the form
\begin{eqnarray}
0 & =  &
\frac{n-2}{2{\overline {r}}}
\left[\frac {d{\overline {H}}}{d{\overline {r}}}
-\frac{n-3}{{\overline {r}}}(k-{\overline {H}})\right]
-\frac{1}{2}{\overline {H}}
\left( \frac {d\Phi}{d{\overline {r}}}\right)^{2}
\nonumber \\ & &
+U(\Phi)+\Lambda ;
\nonumber \\
0 & = &
\frac{n-2}{{\overline {r}}}\frac {d{\overline {\delta}}}{d{\overline {r}}}
-\left(\frac {d\Phi}{d{\overline {r}}} \right) ^2 ;
\nonumber
\\
0 & = &
{\overline {H}} \frac {d^{2}\Phi}{d{\overline {r}}^{2}}
+\frac {d\Phi}{d{\overline {r}}}\left(
{\overline {H}}\frac {d{\overline {\delta}}}{d{\overline {r}}}
+\frac {d{\overline {H}}}{d{\overline {r}}}
+{\overline {H}}\frac{n-2}{{\overline {r}}}\right)
\nonumber \\ & &
 -\frac{dU(\Phi)}{d \Phi} .
\label{eq:conffieldeqs}
\end{eqnarray}

\section{Non-existence of solutions}
\label{sec:no-hair}

In this section we generalize the simple results in \cite{ew2005}
to higher dimensions and solitons.
As in \cite{ew2005}, we emphasize that our method can only be used to show non-existence
of solutions for which the scalar field remains regular everywhere, including at the
origin, infinity and on any event or cosmological horizon.
Therefore our results are evaded by the BBMB black hole \cite{BBMB} as in that
case the scalar field diverges on the event horizon.
We begin with non-existence results which can be proven simply from the scalar
field equation (\ref{eq:scalarfieldeq})
before turning to results where we need to use the conformal transformation of
section \ref{sec:conftrans}.

\subsection{Simple non-existence results}
\label{sec:simple}

Starting with the scalar field equation (\ref{eq:fieldeqs}),
we multiply both sides by $\phi r^{n-2} e^\delta$
and integrate from $r = x$ to $r = y$,
where $x = 0$ for soliton solutions
and $x = r_{h}$, the event horizon, for black hole solutions;
and where $y = r_c$, the radius of the cosmological horizon, for $\Lambda > 0$
and $y = \infty$ for $\Lambda \le 0$.
This gives the equation
\begin{eqnarray}
0 & = & \int_x^y dr \left[\xi R \phi^2 r^{n-2} e^{\delta }
- \phi \left(H \phi' r^{n-2} e^{\delta } \right)' \right]
\nonumber \\
& = &
-\left[\phi H \phi' r^{n-2} e^{\delta } \right]_x^y
\nonumber \\ & &
+ \int_x^y dr \,\, r^{n-2} e^{\delta } \left(\xi R \phi^2 + H \phi'^2\right),
\label{eq:nohair}
\end{eqnarray}
where we have integrated by parts in the second line.

We now need to consider carefully the behaviour of the boundary term
in (\ref{eq:nohair}) at each of the four possible boundary points ($r=0$, $r=r_{h}$,
$r=r_{c}$ and $r=\infty $),
using the boundary conditions derived in Sec.~\ref{sec:boundary}:
\begin{description}
\item[{\bmath {r=0}}:]
At the origin, from (\ref{eq:rto0}),
 we have $H$, $\phi $ and $\delta $ all $O(1)$, while $\phi ' = O(r)$.
 Therefore the boundary term is $O(r^{n-1})$ and vanishes as $r\rightarrow 0$.
\item[{\bmath {r=r_{h}}} and {\bmath {r=r_{c}}}:]
Since $H=0$ at either an event or cosmological horizon, the boundary term vanishes
providing that all the other variables remain finite there.
\item[{\bmath {r=\infty }}:]
These are the most complicated boundary conditions and we need to consider the
asymptotically flat and asymptotically adS cases separately.

Firstly, if $\Lambda = 0$, from (\ref{eq:afinfinity}) we have
$\phi =O(r^{-(n-3)})$ and $\phi '=O(r^{-(n-2)})$, with $H,e^{\delta } \rightarrow 1$,
so that the boundary term is $O(r^{-(n-3)})$ and vanishes as $r\rightarrow \infty $.

Secondly, if $\Lambda < 0$, we have $H=O(r^{2})$, and, from (\ref{eq:phiinf}),
we have $\phi =O(r^{-p})$, $\phi '=O(r^{-p-1})$, so the boundary term is
$O(r^{-2p+n-1})=O(r^{-q})$ where $q$ is given by (\ref{eq:q}).
Following the discussion in Sec.~\ref{sec:boundary}, the boundary term
therefore vanishes only when $\xi <0$ in this case.
\end{description}

Assuming that the boundary term vanishes (as will be the case in the rest of this
section), substituting in for the form of the Ricci scalar using (\ref{eq:ricciscalar})
we can write (\ref{eq:nohair}) as:
\begin{equation}
0 = \int_x^y dr \,\, r^{n-2} e^\delta \left[ \frac{\mathcal{F}}{\mathcal{G}} \right],
\label{eq:simpleintegral}
\end{equation}
where
\begin{eqnarray}
{\mathcal{F}} & = & H \phi'^2 \left(\frac{n}{2} - 1 \right) + n \Lambda \xi \phi^2 ;
\nonumber \\
{\mathcal{G}} & = & 2 \xi \phi^2 \left(n - 1 \right) \left( \xi - \xi_c\right)
+ \left( \frac{n}{2} - 1 \right) .
\label{eq:FandG}
\end{eqnarray}
We now apply (\ref{eq:simpleintegral}) to various cases where we have empty spaces in
Tab.~\ref{tab:PreviousKnowledgeForV0}.
In many of these cases the argument is a simple generalization of the corresponding
result in \cite{ew2005}.

\subsubsection{$\Lambda < 0$ and $\xi < 0$}
\label{subsec:Lnegxineg}

In this case the boundary term at infinity vanishes, as well as at the origin or
black hole event horizon.
For these parameter values the function $\mathcal{G} >0$ and finite everywhere
inside the integration range.
It is straightforward to see that $\mathcal{F}$ is a sum of positive terms,
and therefore the only way the integral (\ref{eq:simpleintegral})
can vanish is if $\mathcal{F} \equiv 0$ everywhere inside the integration range.
This is only possible if $\phi \equiv 0$ everywhere.

\subsubsection{$\Lambda > 0$ and $\xi > \xi_c$}
\label{subsec:Lposxibigxic}

The non-existence proof in this case follows the argument in Sec.~\ref{subsec:Lnegxineg},
except that here we consider only the region inside the cosmological horizon.
Again $\mathcal{G} >0$ and finite everywhere within the integration range.
Also, the function $\mathcal{F}$ is again a
sum of positive terms so the only solution is again $\phi \equiv 0$.

\subsubsection{$\Lambda > 0$ and $0 <\xi < \xi_c$}
\label{subsec:Lposxibet0xic}

Here the argument is slightly more complicated as $\mathcal{G}$
could have a zero somewhere in the integration range.
However, at such a point the Ricci scalar curvature would be infinite unless
$\mathcal{F}$ also has a zero at that point.
Now ${\mathcal {F}}$ is a sum of positive terms,
so if ${\mathcal {F}}=0$ at a point, each term must be zero at that point,
meaning that $\phi =0 $ at that point.
Substituting $\phi =0$ into ${\mathcal {G}}$ gives ${\mathcal {G}} \neq 0$,
which is a contradiction, so ${\mathcal {G}}$ cannot have any zeros in the
region of integration.
As  $\mathcal{F}$ is always positive and $\mathcal{G}$ is of constant sign,
the only way the integral (\ref{eq:simpleintegral}) can be zero is
if  $\mathcal{F} \equiv 0$ everywhere,
which again gives us the trivial solution $\phi \equiv 0$.

\vspace{0.5cm}

\subsubsection{$\Lambda = 0$, $\xi < 0$ and $\xi > \xi_c$}
\label{subsec:Lzeroxinegorbigxic}

When $\xi <0$ or $\xi > \xi _{c}$ it is the case that ${\mathcal {G}}>0$.
With $\Lambda = 0$, the function ${\mathcal {F}}$ simplifies to give a single positive
term.
Therefore, as in previous subsections, $\phi' \equiv 0$,
and the condition that $\phi \rightarrow 0 $ as $r \rightarrow \infty $ means that
$\phi \equiv 0 $ everywhere.
These results are in agreement with those derived in \cite{saa} by using
the conformal transformation of Sec.~\ref{sec:conftrans}.

\vspace{-0.7cm}

\subsubsection{$\xi = 0$}
\label{subsec:xieqzero}

The minimally coupled case is very straightforward:
the function $\mathcal{G} = \frac{n}{2} - 1 > 0$ everywhere.
Once again ${\mathcal {F}}$ reduces to a single positive term,
and as in Sec.~\ref{subsec:Lzeroxinegorbigxic},
it must be the case that $\phi ' \equiv 0$, and hence $\phi \equiv 0$.

\vspace{-0.7cm}

\subsubsection{$\xi = \xi_c$ and $\Lambda \ge 0$}
\label{subsec:xieqxicLpos}

In the conformally coupled case, the function $\mathcal{G} = \frac{n}{2} - 1 > 0$,
exactly as in Sec.~\ref{subsec:xieqzero}.
The same argument holds
about $\mathcal{F}$ being sum of positive terms
so only possible solution is the trivial one
where $\phi \equiv 0$.

\begin{widetext}
\subsubsection{Summary}
\label{subsec:summary}

At this stage it is useful to update
Tab.~\ref{tab:PreviousKnowledgeForV0} to incorporate our new results.
\begin{table}[h]
\begin{tabular}[c]{|c||c|c|c|c|c|}
\hline
$n>4$ & $\xi < 0$ & $\xi = 0$ & $0 < \xi < \xi_c$ & $\xi = \xi_c$ & $\xi > \xi_c$ \\
\hline
\hline
$\Lambda > 0$
& see Sec.~\ref{sec:confnohair}
& {\bf {no solutions}}
& {\bf {no solutions}}
& {\bf {no solutions}}
& {\bf {no solutions}}
\\
\hline
$\Lambda = 0$
& {\bf {no solutions}}
& {\bf {no solutions}}
& no solutions \cite{saa}
& {\bf {no solutions}}
& {\bf {no solutions}}
\\
\hline
$\Lambda < 0$
& {\bf {no solutions}}
& {\bf {no solutions}}
&
& stable solutions \cite{raduew}
&
\\
\hline
\end{tabular}
\caption{Summary of existence and non-existence of soliton and black hole
solutions in more than four space-time dimensions.}
\label{tab:KnowledgeForV0v2}
\end{table}
In Tab.~\ref{tab:KnowledgeForV0v2}, we have highlighted new results in bold, and
emphasize that our results rule out both soliton and black hole solutions, including
topological black holes.
We are not able to use our simple approach in this section to say anything about
the case $\Lambda = 0$, $0<\xi <\xi _{c}$, which was considered in \cite{saa}
using the conformal transformation of Sec.~\ref{sec:conftrans}.
We are so far unable to say anything about the cases $\Lambda >0,\xi <0$; $\Lambda <0,
0<\xi <\xi _{c}$ and $\Lambda <0, \xi > \xi_{c}$.
The first of these will be dealt with in the next section.
\end{widetext}

\subsection{Non-existence result for $\Lambda >0$ and $\xi <0$}
\label{sec:confnohair}

The non-existence of black hole solutions in this case was shown in four
space-time dimensions in \cite{ew2005}.
That result extends trivially to higher-dimensions and solitons, so we only
provide a very brief outline.

Since $\xi <0$, the conformal transformation (\ref{eq:conftransgmunu}) is
valid providing all field variables remain finite.
The transformed scalar field equation (\ref{eq:conffieldeqs}) gives, at
the {\em {cosmological}} horizon $r=r_{c}$,
\begin{equation}
\frac{d\Phi}{d{\overline{r}}}
\frac{d{\overline{H}}}{d{\overline{r}}} =
\frac{dU(\Phi)}{d\Phi}.
\label{eq:transfieldeqhoriz}
\end{equation}
Now ${\overline{H}} >0$ inside the cosmological horizon,
and ${\overline{H}} <0$ outside the cosmological horizon,
so $\frac{d{\overline{H}}}{d{\overline{r}}}<0$ on the cosmological horizon.
We then have two cases:
$\frac{d\Phi}{d{\overline{r}}}>0$,
$\frac{dU(\Phi)}{d\Phi}<0$ and vice versa.

The argument is the same for both cases so we only consider
$\frac{d\Phi}{d{\overline{r}}}>0$,
$\frac{dU(\Phi)}{d\Phi}<0$.
With $\Lambda >0$ and $\xi < 0$, the potential $U(\Phi )$ is everywhere
negative and has a maximum at $\Phi =0$ (see Fig.~\ref{fig:UPhi}).
Therefore $\Phi $ is positive and increasing on the horizon.
In order to satisfy the boundary condition on the original scalar field,
namely $\phi \rightarrow 0$ at infinity,
it must be the case that $\Phi \rightarrow 0$ at infinity, and hence
$\Phi $ must have a maximum somewhere outside the cosmological horizon.
At such a maximum $\frac{d\Phi}{d{\overline{r}}}=0$
and the field equation (\ref{eq:conffieldeqs}) gives:
\begin{equation}
{\overline{H}} \frac{d^2\Phi}{d{\overline{r}^2}}
= \frac{dU(\Phi)}{d\Phi}<0.
\label{eq:transfieldeqmaxPhi}
\end{equation}
However, since we are outside the cosmological horizon, it is the case that
${\overline {H}}<0$ and so equation (\ref{eq:transfieldeqmaxPhi})
gives $\frac{d^2\Phi}{d{\overline{r}^2}}>0$ which is a contraction.
Therefore there can be no non-trivial solutions in this case.

The conformal transformation can also be applied to the case $\Lambda =0$,
$0<\xi <\xi _{c}$, when the transformed potential
$U(\Phi )=0$ (\ref{eq:conftransfinalU}) \cite{saa}.

\section{Non-trivial solutions}
\label{sec:nontrivial}

In the previous section, we were able to rule out non-trivial
higher-dimensional soliton or
black hole solutions of the field equations (\ref{eq:fieldeqs}) except
when $\Lambda <0$ and $0<\xi < \xi _{c}$ or $\xi > \xi _{c}$,
solutions with $\Lambda <0$ and $\xi = \xi _{c}$ having already been
found \cite{raduew}.
It is no surprise that we find non-trivial solutions for these other values of
$\xi $.
In four dimensions, spherically symmetric black hole solutions were found
in \cite{ew2005}.
We find solutions which generalize these to higher dimensions, as well as
solitons and topological black holes in four and more space-time dimensions.
In this section we first discuss the properties of the numerical solutions,
before studying their thermodynamics and stability.

\subsection{Numerical solutions}
\label{sec:numerics}

Using a standard ordinary differential equation solver, we integrate the
field equations (\ref{eq:fieldeqs}), starting close to either the origin
or black hole event horizon, as applicable, and integrating towards
$r \rightarrow \infty $.
Since the field equations (\ref{eq:fieldeqs}) are invariant under the
transformation $\phi \rightarrow -\phi $, we only consider positive
values of $\phi $ at the origin or event horizon.
Without loss of generality, we set $\Lambda = -(n-1)(n-2)/2$, so that
the length scale set by the cosmological constant,
$\ell ^{2}= - (n-2)(n-1)/(2\Lambda )=1$ (this can be achieved by a rescaling
of the co-ordinates and metric function $m$).

We find soliton, spherically symmetric ($k=1$) black holes and topological
black holes ($k\neq 1$) for any number of space-time dimensions, $n\ge 4$,
and any value of the coupling constant $\xi >0$.
Our numerical work indicates that solutions exist for any value of $\phi $ at the
origin or event horizon such that $0<\phi <\xi ^{-1/2}$.

\begin{figure}
\includegraphics[width=6.5cm,angle=270]{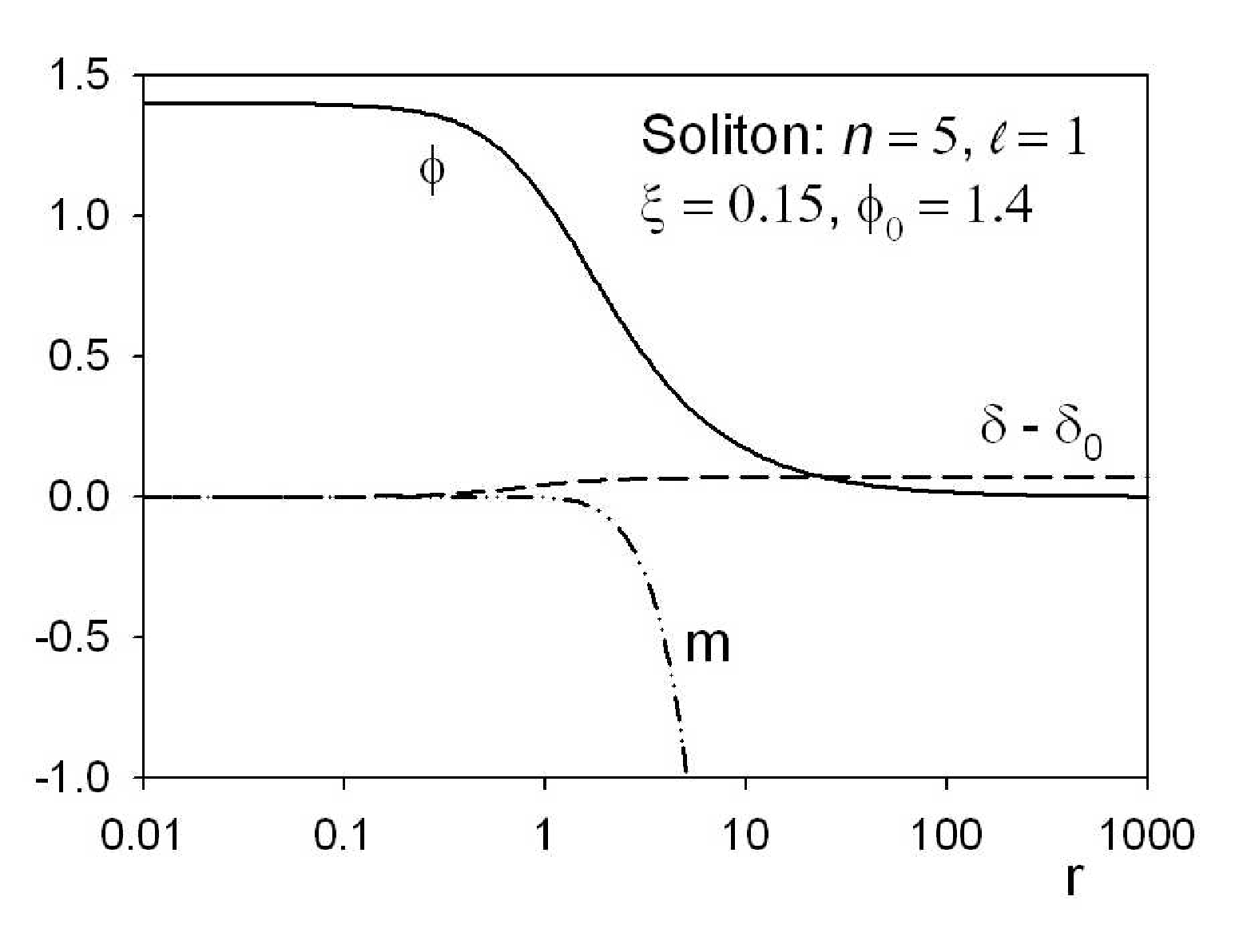}
\caption{
Example of a five-dimensional soliton solution with $\ell = 1$, $\xi = 0.15$
and $\phi _{0}=1.4$.
In this case the scalar field $\phi $ is monotonic and has no zeros.
}
\label{fig:sol5d}
\end{figure}

\begin{figure}
\includegraphics[width=6.5cm,angle=270]{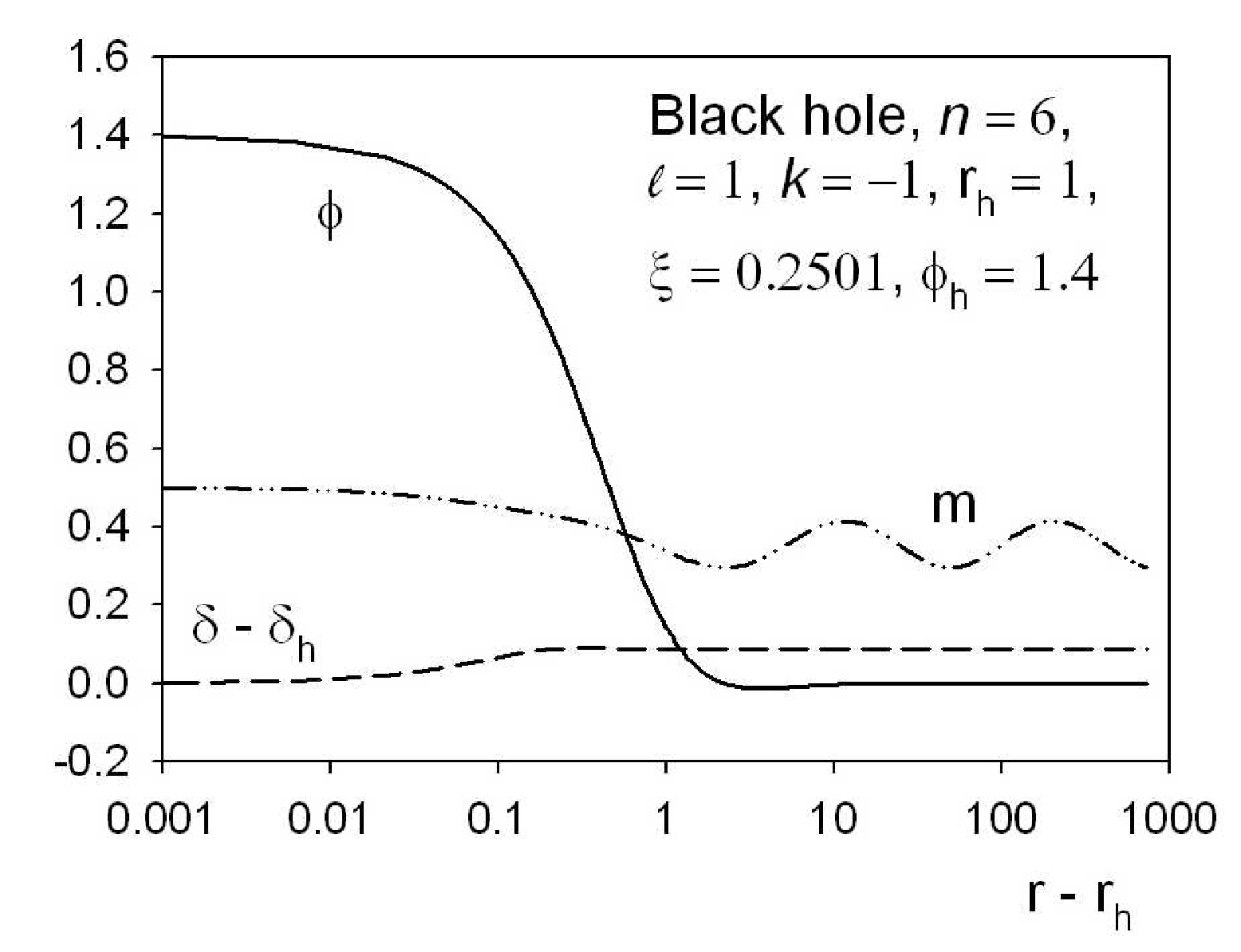}
\caption{
Example of a six-dimensional topological black hole solution
with $k=-1$, $\ell = 1$, $\xi = 0.2501$
and $\phi _{0}=1.4$.
In this case we find that the scalar field $\phi $ oscillates
as it tends to zero as $r\rightarrow \infty $, although the magnitude of the
oscillations is too small to see on the graph.
}
\label{fig:bh6d}
\end{figure}

Two typical solutions are illustrated in Figs.~\ref{fig:sol5d} and \ref{fig:bh6d}.
In Fig.~\ref{fig:sol5d}, a typical five-dimensional soliton solution is shown,
for $\xi = 0.15$. This value of $\xi $ is less than $(n-1)/(4n)=0.20$, and, as predicted
in Sec.~\ref{sec:boundary}, the scalar field $\phi $ is monotonic and has no zeros.
In Fig.~\ref{fig:bh6d} we plot a typical six-dimensional black hole solution with
$\xi = 0.2501$.  Since this value of $\xi $ is greater than $(n-1)/(4n) = 5/24$,
in this case we find that the scalar field $\phi $ oscillates as it tends to zero
as $r\rightarrow \infty $, as predicted in Sec.~\ref{sec:boundary}, although the
magnitude of the oscillations is too small to see in Fig.~\ref{fig:bh6d}.

\begin{figure}
\includegraphics[width=6.5cm,angle=270]{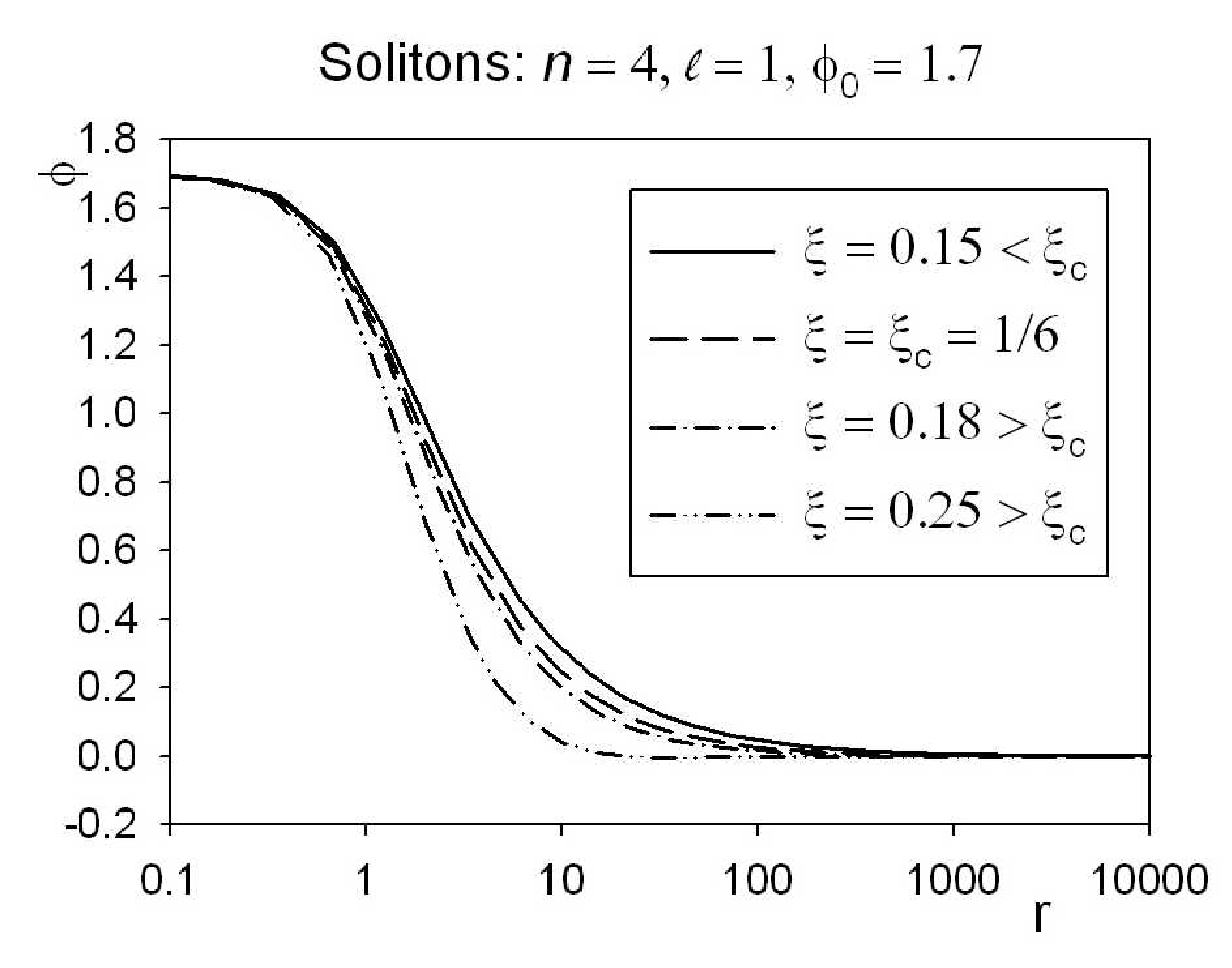}
\caption{
The effect of varying $\xi $ is shown for four-dimensional soliton solutions
with $\ell = 1$ and $\phi _{0} = 1.7$.
As discussed in Sec.~\ref{sec:boundary},
for $\xi < (n-1)/(4n)$, the scalar field $\phi $ monotonically decreases to zero
as $r\rightarrow \infty $ and has no zeros, but for $\xi > (n-1)/(4n)$,
the scalar field oscillates as it tends to zero, although the magnitude of
the oscillations is very small.
Apart from this, varying $\xi $ does not have a great effect on the form
of the scalar field $\phi $.
For higher dimensions, the effect on $\phi $ of varying $\xi $ is even less.
}
\label{fig:sol4dvaryxiphi}
\end{figure}

This change in the behaviour of the scalar field as $r\rightarrow \infty $
on varying the coupling constant $\xi $ is the same as found for the four-dimensional
solutions \cite{ew2005}, although typically the oscillations in $\phi $ for
$\xi > (n-1)/(4n)$ have only a small magnitude.
Apart from these oscillations, the values of the function $\phi $ do not seem to
change very much as $\xi $ varies, as shown in Fig.~\ref{fig:sol4dvaryxiphi}.

\begin{figure}
\includegraphics[width=8cm]{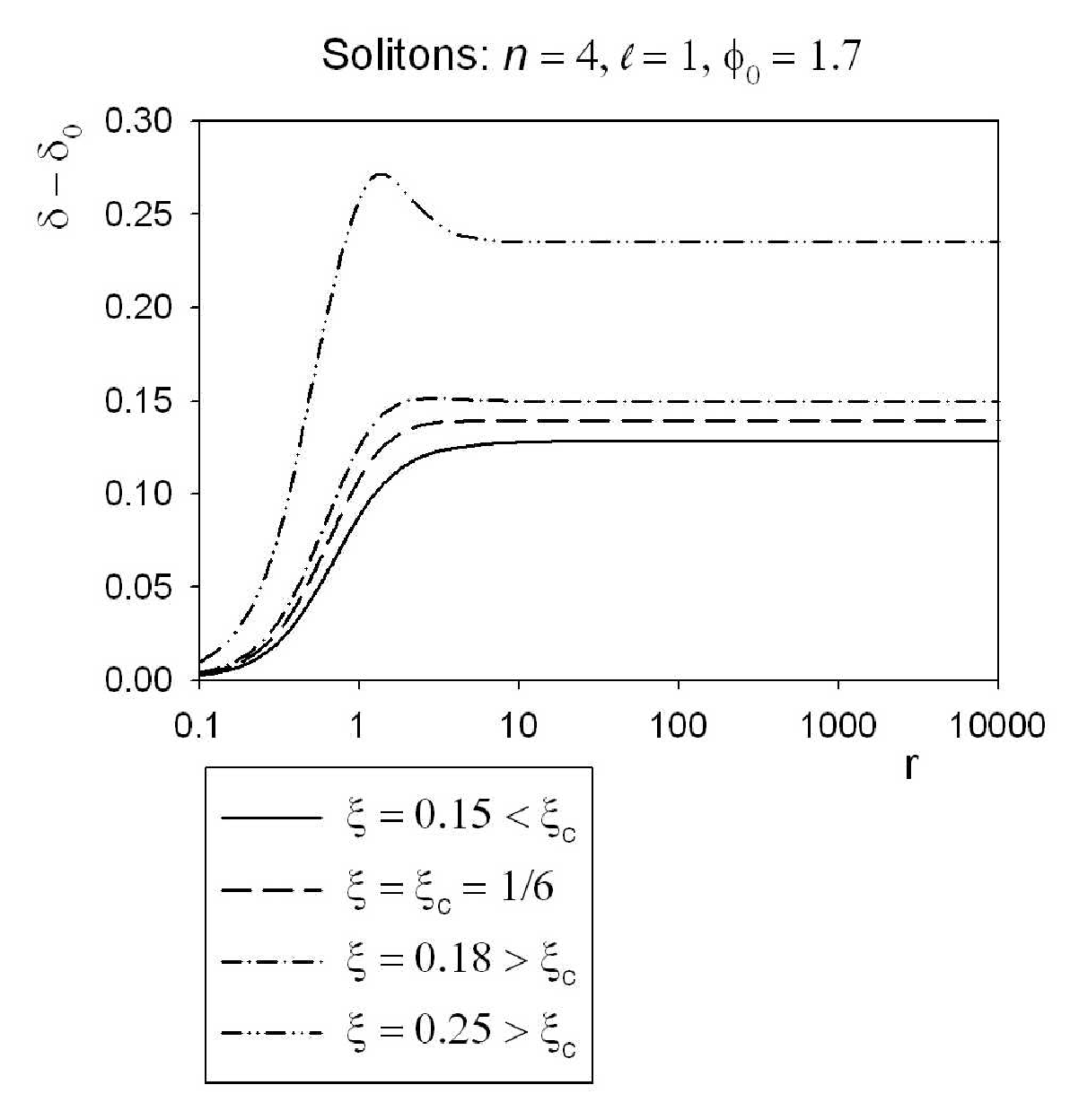}
\caption{
The effect on the metric function $\delta $  of varying $\xi $
is shown for four-dimensional soliton solutions
with $\ell = 1$ and $\phi _{0} = 1.7$.
The function $\delta $ has the limit $\delta \rightarrow 0$ as $r\rightarrow \infty $,
so we have plotted $\delta - \delta _{0}$ in order to make the
behaviour of the functions easier to visualize.
}
\label{fig:sol4dvaryxidelta}
\end{figure}
\begin{figure}
\includegraphics[width=8cm]{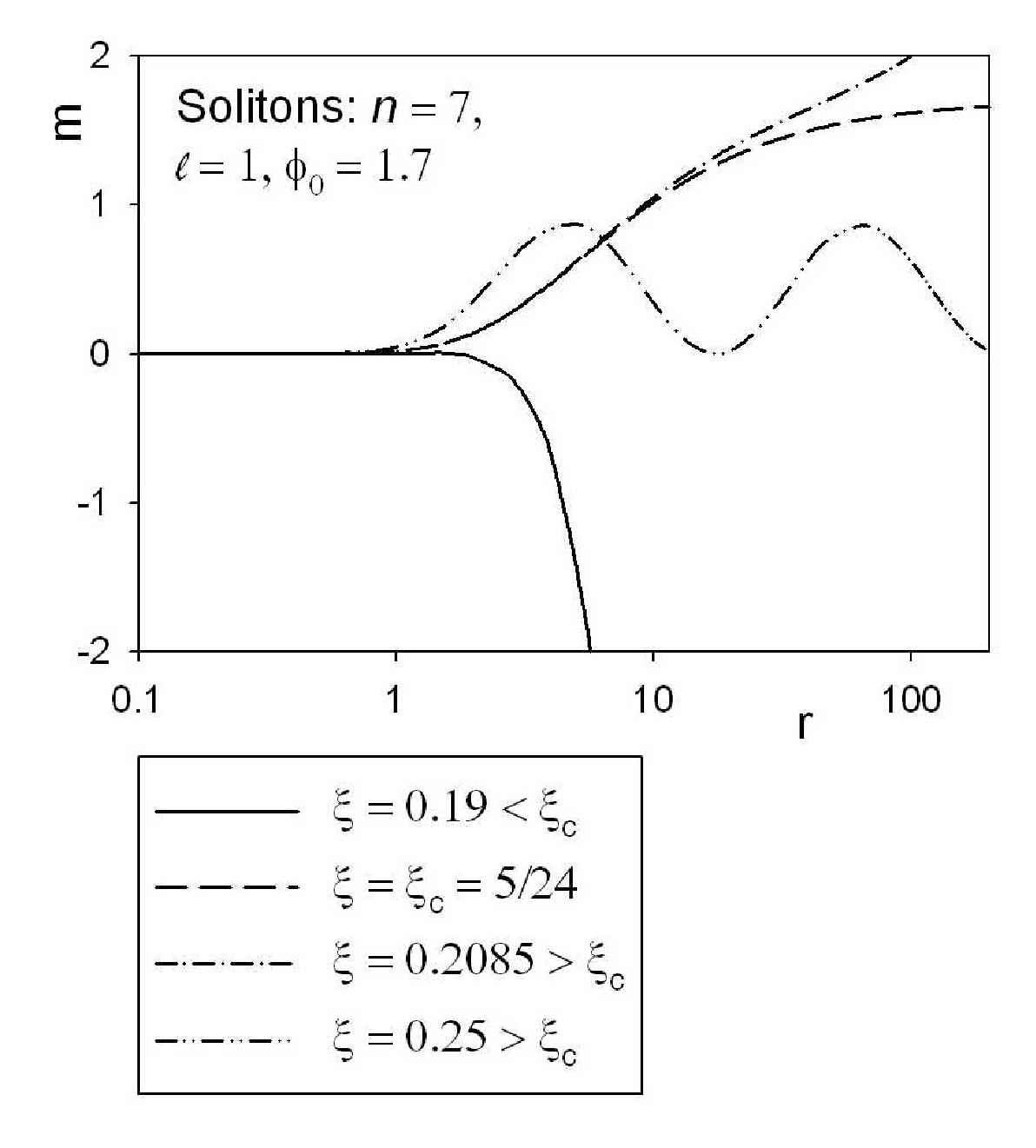}
\caption{
The effect on the metric function $m$ of varying $\xi $
is shown for seven-dimensional soliton solutions
with $\ell = 1$ and $\phi _{0} = 1.7$.
}
\label{fig:sol7dvaryxim}
\end{figure}
\begin{figure}
\includegraphics[width=8cm]{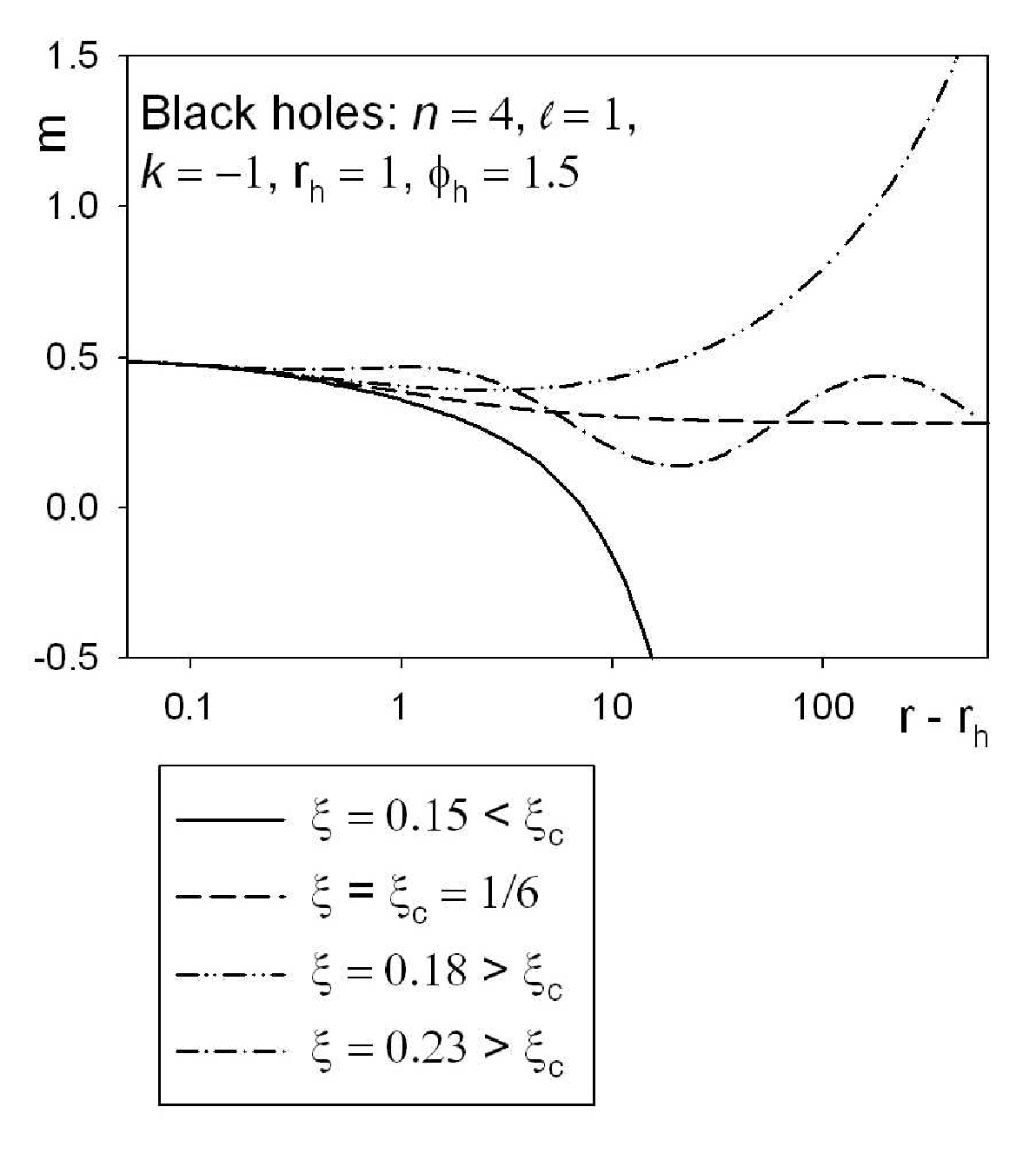}
\caption{
The effect on the metric function $m$ of varying $\xi $ is shown
for four-dimensional black hole solutions with $\ell =1$, $r_{h}=1$,
$k=-1$ and $\phi _{h}=1.5$.
}
\label{fig:mvaryxik-1}
\end{figure}
Varying the coupling constant $\xi $ has a much more significant effect on the
metric functions $\delta $ and $m$, as shown in
Figs.~\ref{fig:sol4dvaryxidelta}--\ref{fig:mvaryxik-1}.
The effect on the metric function $\delta $ is shown in
Fig.~\ref{fig:sol4dvaryxidelta} for some four-dimensional soliton solutions.
In four dimensions, the value of $\xi $ at which the oscillatory behaviour of
the scalar field begins is $\xi = 3/16=0.1875$.
It can be seen from Fig.~\ref{fig:sol4dvaryxidelta} that for values of
$\xi $ below 0.1875, the function $\delta $ is monotonically increasing with $r$,
although its value at the origin is decreasing as $\xi $ increases (recall that
$\delta \rightarrow 0 $ as $r\rightarrow \infty $, although in
Fig.~\ref{fig:sol4dvaryxidelta} we have plotted $\delta -\delta _{0}$ rather than $\delta $
to make the graph easier to read).
However, for values of $\xi $ greater than 0.1875, the function $\delta $ has
a maximum before decreasing monotonically to its value at infinity.

As discussed in Sec.~\ref{sec:boundary}, the effect of $\xi $ on the behaviour
of the metric function $m$ is even more dramatic, and shown in
Figs.~\ref{fig:sol7dvaryxim} and \ref{fig:mvaryxik-1}.
When $\xi < \xi _{c}= (n-2)/[4(n-1)]$, the metric function $m$ diverges to $-\infty $
as $r\rightarrow \infty $.
Only for $\xi =\xi _{c}$ does the metric function $m$ converge as $r \rightarrow
\infty $, as observed in \cite{raduew}.
The function $m$ diverges to $+\infty $ as $r\rightarrow \infty $ for values of
$\xi$ between $(n-2)/[4(n-1)]$ and $(n-1)/(4n)$, a shrinking interval as $n$ increases.
For $\xi > (n-1)/(4n)$, the metric function $m$ oscillates
about a non-zero value as $r\rightarrow \infty $.

\begin{figure}
\includegraphics[width=6.0cm,angle=270]{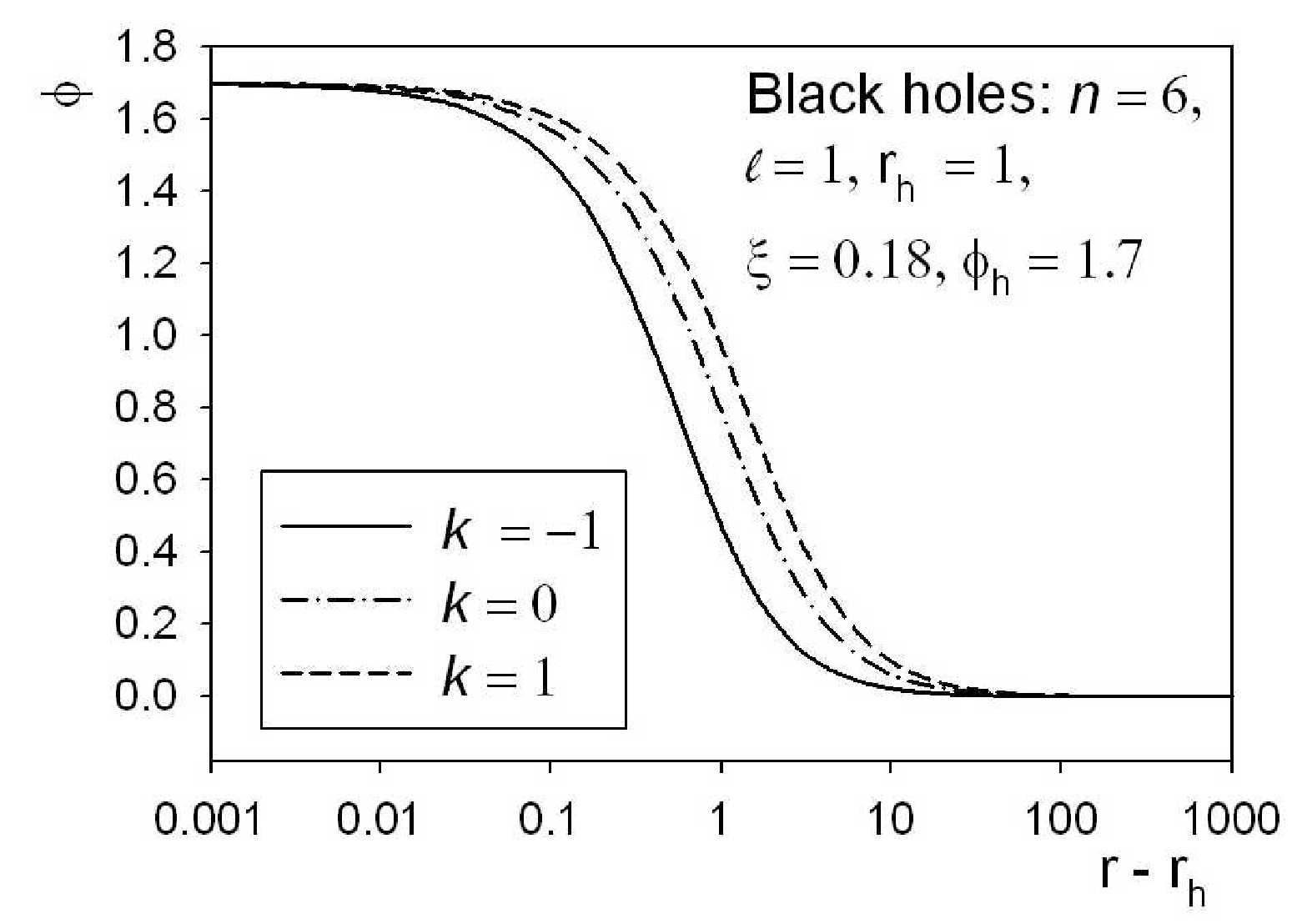}
\caption{
The effect on the scalar field $\phi $ of varying the constant $k$ is shown
for some six-dimensional black holes with $\ell = 1$, $r_{h}=1$, $\xi = 0.18$
and $\phi _{h} = 1.7$.
The functions $\phi $ for the different values of $k$ do not differ significantly.
}
\label{fig:bh6dvarykphi}
\end{figure}
\begin{figure}
\includegraphics[width=6.5cm,angle=270]{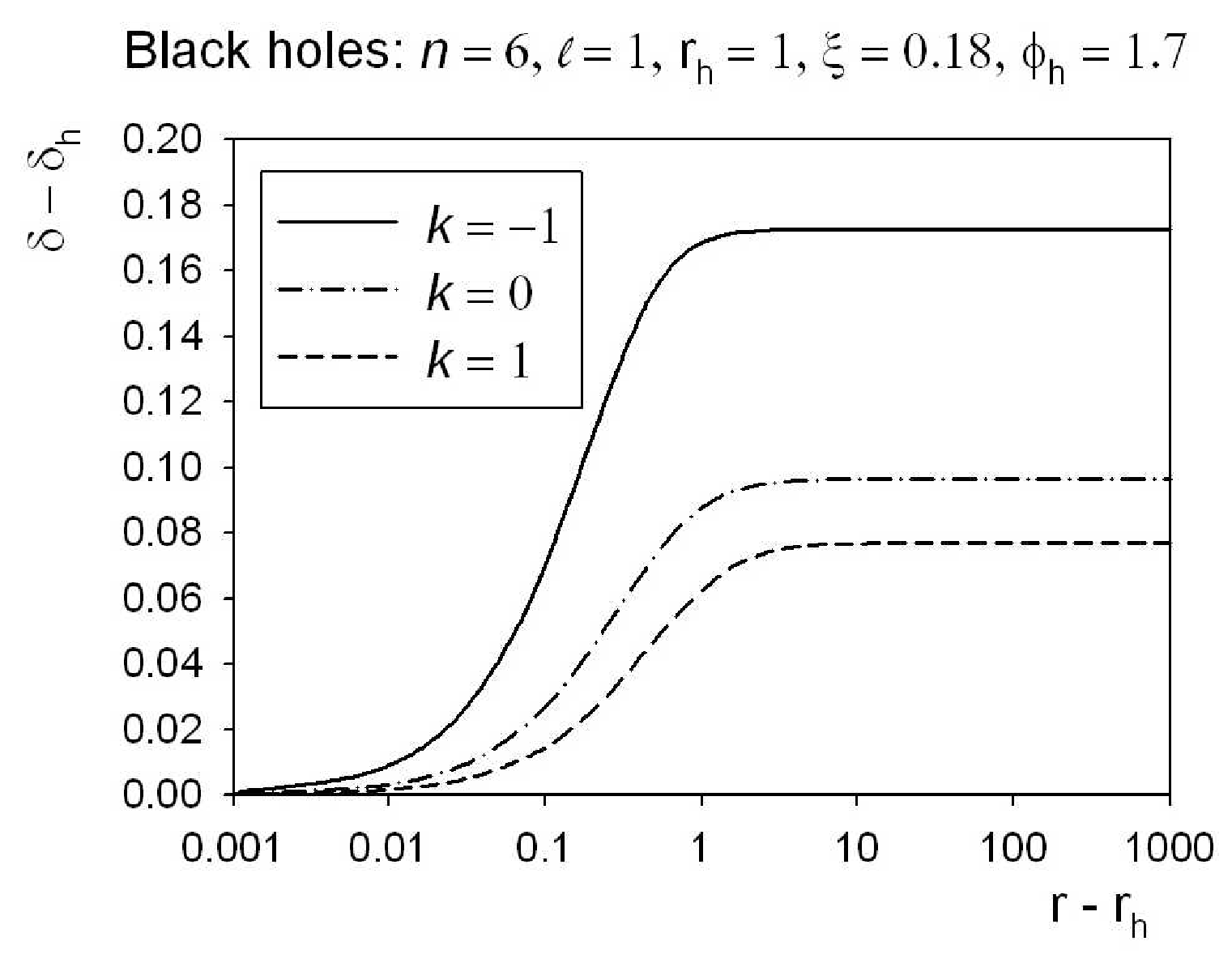}
\caption{
The effect on the metric function $\delta $ of varying the constant $k$ is shown
for the same black hole solutions as in Fig.~\ref{fig:bh6dvarykphi}.
The function $\delta $ has the limit $\delta \rightarrow 0$ as $r\rightarrow \infty $,
so we have plotted $\delta - \delta _{h}$ in order to make the
behaviour of the functions easier to visualize.
}
\label{fig:bh6dvarykdelta}
\end{figure}
For black hole solutions, there is an additional parameter to vary, namely the constant
$k\in \{ 0, 1, -1 \} $, which governs the topology of the event horizon.
The effect on the scalar field $\phi $ and metric function $\delta $ of varying $k$
is shown in Figs.~\ref{fig:bh6dvarykphi} and \ref{fig:bh6dvarykdelta} respectively.
Varying the constant $k$ does not significantly change the scalar field $\phi $, as
seen in Fig.~\ref{fig:bh6dvarykphi}, in agreement with the observations in
the conformally coupled case \cite{raduew}.
However, we find more significant differences in the magnitude of the function
$\delta $ in the non-conformally coupled case
than in the conformally coupled case (compare Fig.~\ref{fig:bh6dvarykdelta}
and \cite{raduew}).

\begin{figure}
\includegraphics[width=6.0cm,angle=270]{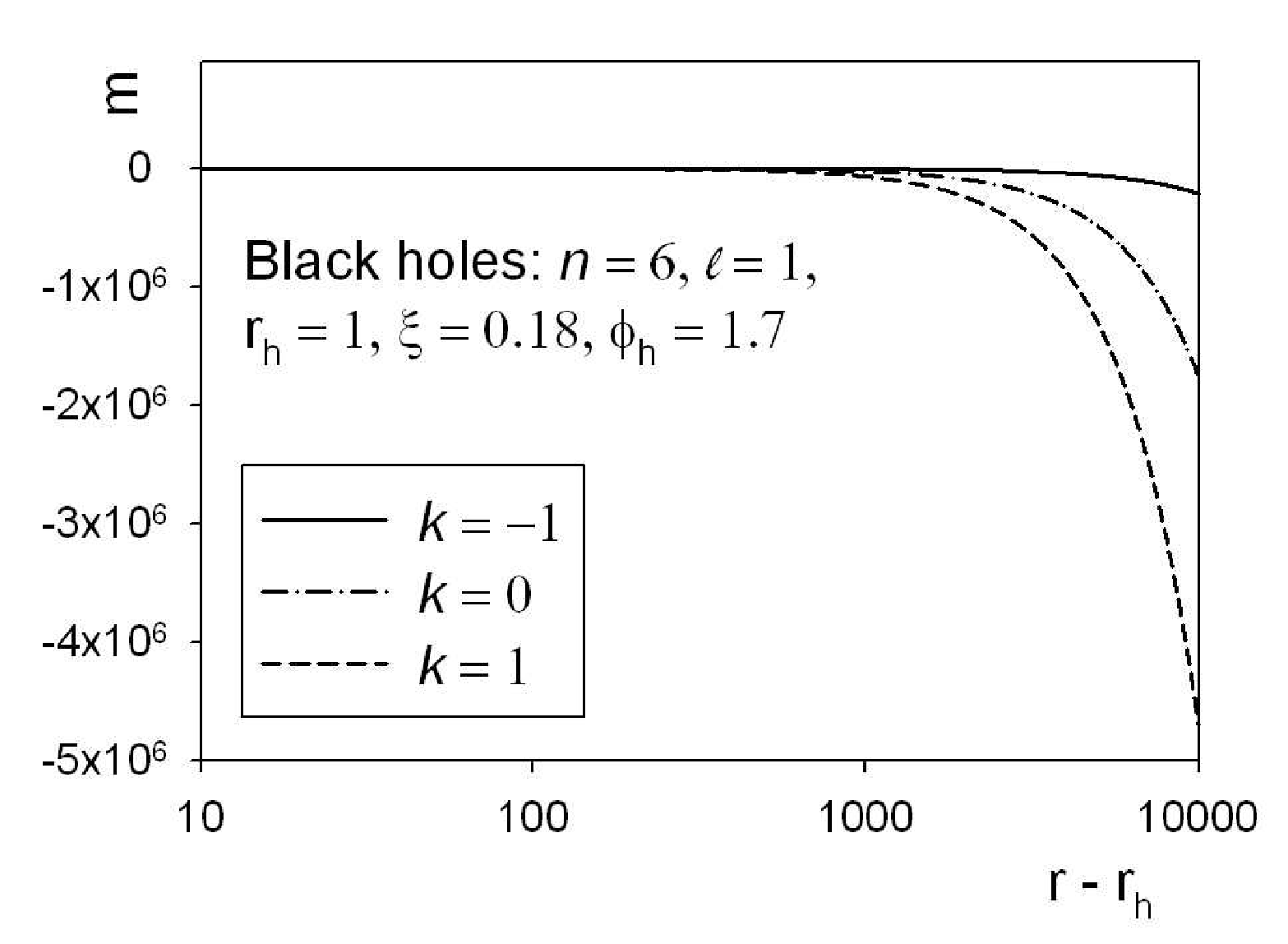}
\caption{
The effect on the metric function $m$ of varying the constant $k$ is shown
for the same black hole solutions as in Fig.~\ref{fig:bh6dvarykphi}.
}
\label{fig:bh6dvarykmbelow}
\end{figure}

\begin{figure}
\includegraphics[width=6.0cm,angle=270]{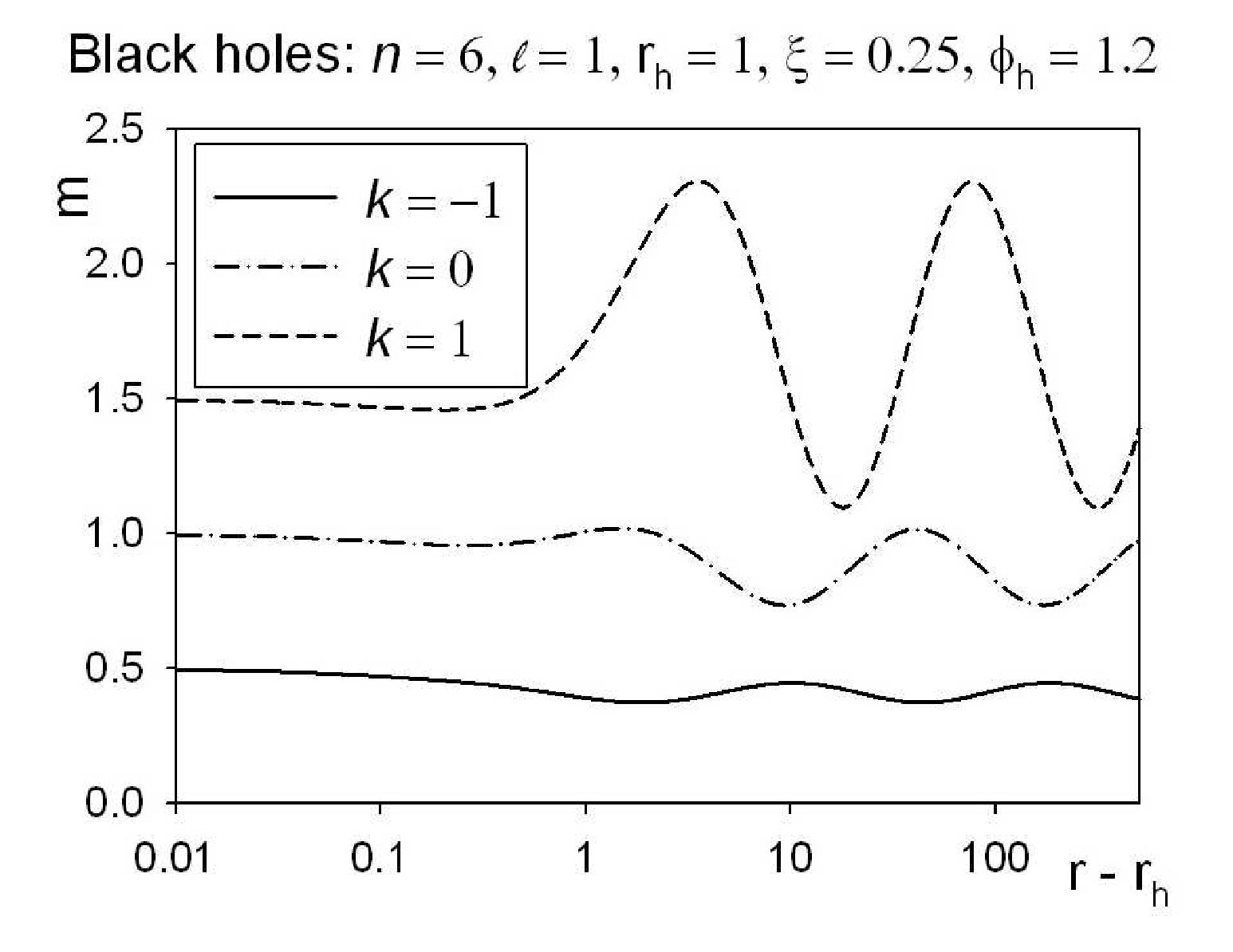}
\caption{
The effect on the metric function $m$ of varying the constant $k$ is shown
for the some six-dimensional black hole solutions
with $\ell = 1$, $r_{h}=1$, $\xi = 0.25$ and $\phi _{h} = 1.2$.
}
\label{fig:bh6dvarykmabove}
\end{figure}

When the scalar field is conformally coupled and the metric function $m$
converges to a finite limit as $r\rightarrow \infty $, it was observed in \cite{raduew}
that varying $k$ had a significant impact on the properties of the metric function
$m$ (see, Fig.~3 in \cite{raduew}).
For non-conformal coupling, when the metric function $m$ either diverges or oscillates
as $r\rightarrow \infty $, we find that varying $k$ does not alter these dominant
behaviours very much, as illustrated in Figs.~\ref{fig:bh6dvarykmbelow} and
\ref{fig:bh6dvarykmabove}.
When the metric function $m$ diverges, as seen in Fig.~\ref{fig:bh6dvarykmbelow},
the rate of divergence of the function $m$ as $r\rightarrow \infty $ does not
change as $k$ varies, but the coefficient $M_{1}$ multiplying the dominant divergent
term in the behaviour of $m$ (\ref{eq:mansatz}) does vary (see
Eq.~(\ref{eq:M1sign}) - the constant $M_{1}$ depends on the constant $c_{1}$
in the expansion of the scalar field $\phi $ at infinity (\ref{eq:phiinf}),
which varies as $k$ varies).
When the metric function $m$ oscillates as $r\rightarrow \infty $, the effect of
varying $k$ can be seen in Fig.~\ref{fig:bh6dvarykmabove}.
In accordance with the analysis of Sec.~\ref{sec:boundary}, it can be seen
that changing $k$ does not affect the period of the oscillations, but does affect
their magnitude and phase.
In particular, the magnitude of the oscillations increases as $k$ increases.

\subsection{Thermodynamics}
\label{sec:thermo}

The thermodynamics of black holes with minimally coupled scalar field hair has
recently attracted attention in relation to the adS/CFT correspondence
\cite{zeng,adSCFThair,SUGRAhairstab}, and a number of authors
have studied the thermodynamics of black holes with a conformally
coupled scalar field \cite{nadalini,thermo}.
Given the difficulties, outlined in Sec.~\ref{sec:boundary}, of defining
a mass for our non-minimally coupled solutions, we do not attempt a full
thermodynamical analysis here.
Instead we just make some brief remarks about the temperature and entropy of
the black holes.

The temperature is given by the usual Hawking formula:
\begin{equation}
T = \frac {1}{4\pi } H'(r_{h}) e^{\delta _{h}};
\label{eq:temperature}
\end{equation}
while the entropy is modified by the non-minimally coupled scalar field \cite{entropy}:
\begin{equation}
S = 2\pi V r_{h}^{n-2} \left[ 1 - \xi \phi _{h} ^{2} \right] ,
\label{eq:entropy}
\end{equation}
where $V$ is the volume of the maximally symmetric space with metric
$d\sigma _{n-2,k}$ (\ref{eq:dsigma}).

\begin{figure}[h]
\includegraphics[width=6.0cm,angle=270]{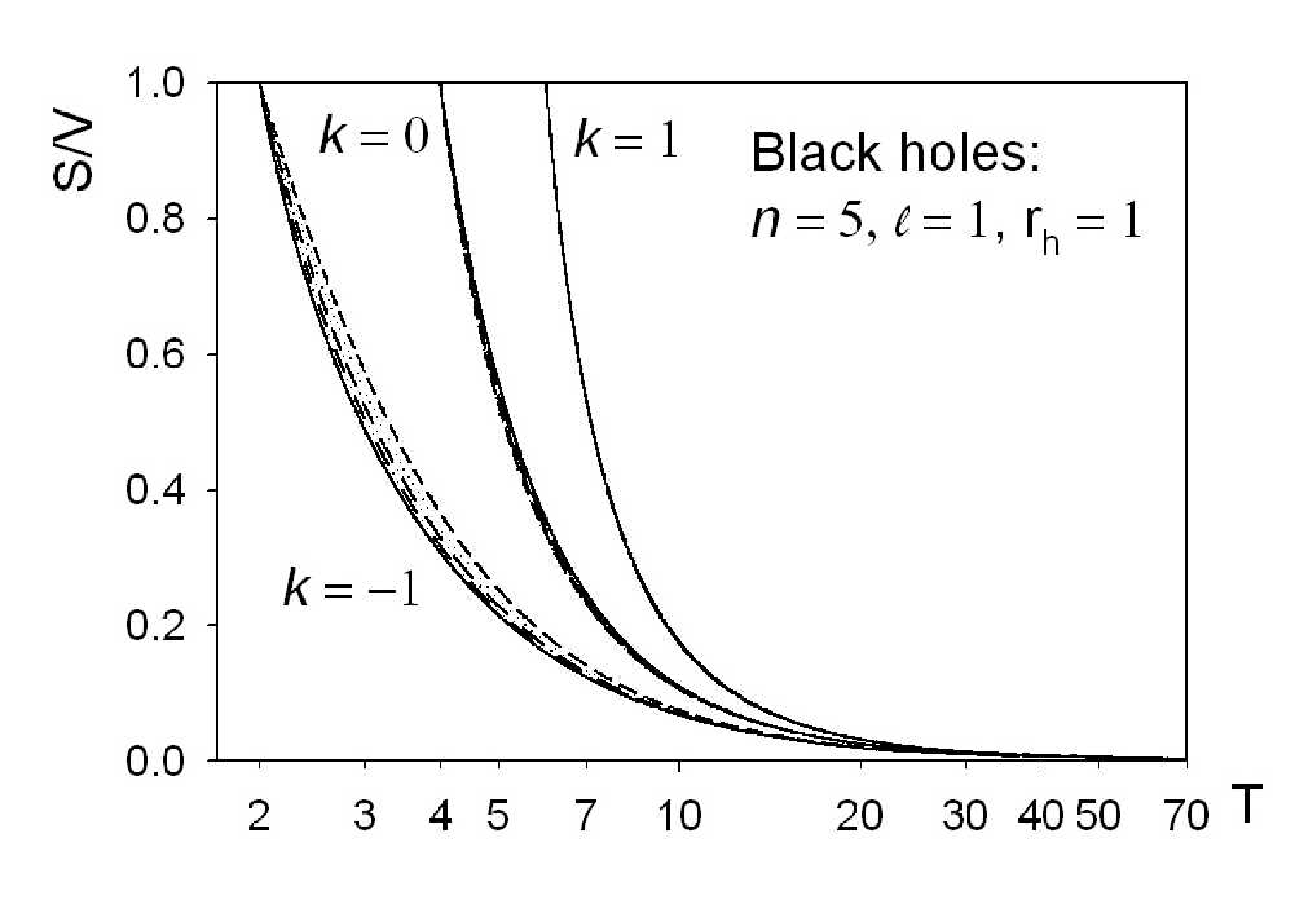}
\caption{
Temperature-entropy relation for some five-dimensional black holes with
fixed horizon radius $r_{h}=1$, and $\ell = 1$.
For each value of $k$, we have plotted curves for five values of $\xi $:
0.15, 0.1875, 0.225, 0.2625 and 0.3, and varied the value of the scalar
field on the horizon $\phi _{h}$ such that both $S$ and $T$ are positive.
}
\label{fig:TSvaryxi}
\end{figure}

In Fig.~\ref{fig:TSvaryxi}, we have plotted the relationship between temperature $T$
and entropy $S/V$ for some five-dimensional black holes with $r_{h}=1$, varying
$k$ and, for each value of $k$, considering five values of the coupling constant
$\xi $.
For each curve, we computed the temperature and entropy for all values of $\phi _{h}$
such that $0<\phi _{h}< 1/{\sqrt {\xi }}$.
It can be seen from Fig.~\ref{fig:TSvaryxi} that varying $k$ and $\xi $ makes little
difference to the relationship between temperature and entropy, particularly for $k=0$
and $k=1$ (in the latter case the curves for different $\xi $ are
indistinguishable).

\begin{widetext}
In Fig.~\ref{fig:TSseparaterh} we fix $k=1$ and $\xi = 1/6$ and consider the effect
of changing the event horizon radius $r_{h}$ on the temperature and entropy of some
five-dimensional black holes.
\begin{figure}[h]
\includegraphics[width=12cm,angle=270]{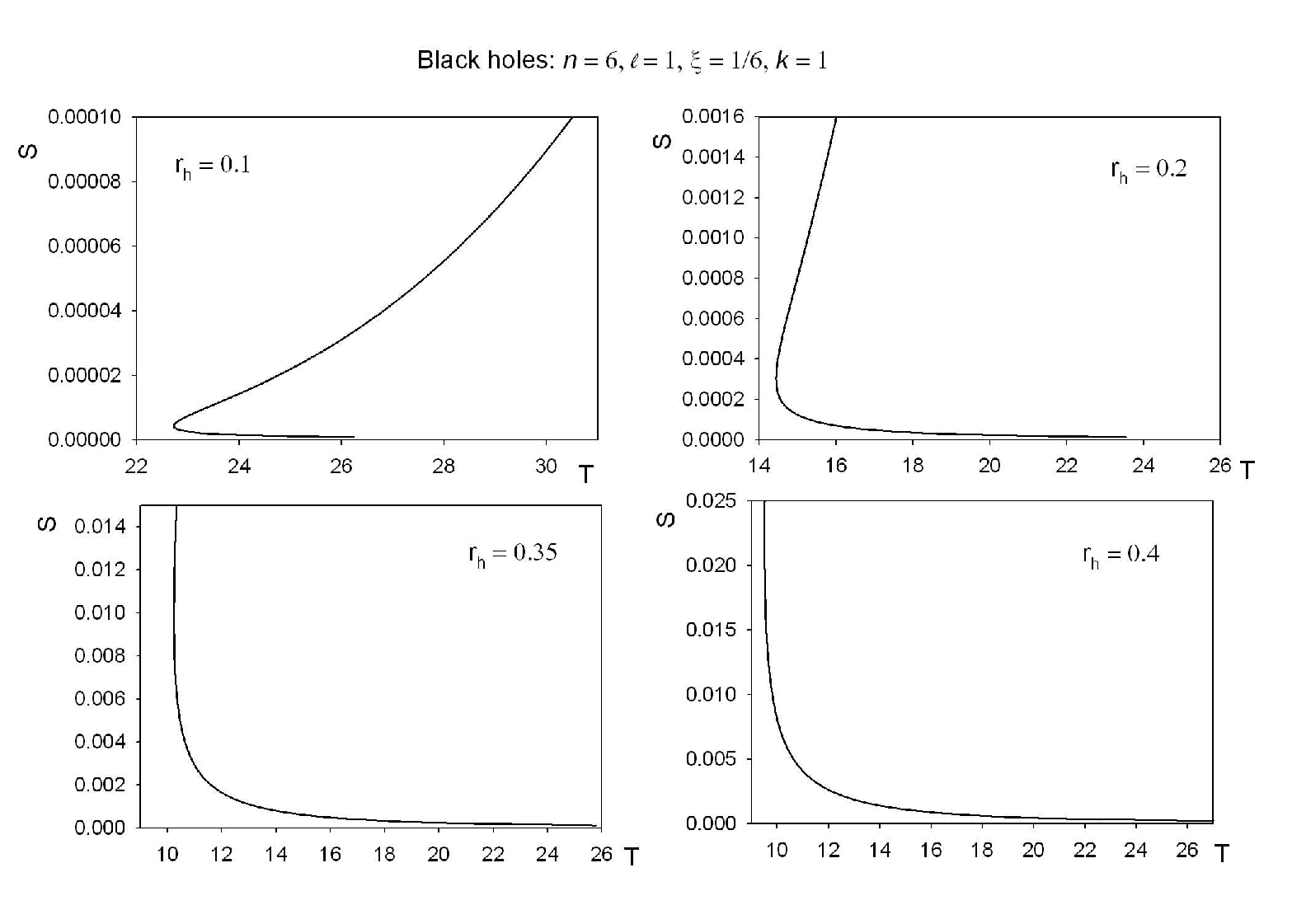}
\caption{Temperature-entropy relations for some values of the event horizon
radius $r_{h}$, for five-dimensional black holes
with $k=1$, $\ell = 1$ and $\xi = 1/6$.
For each value of $r_{h}$ considered, we vary $\phi _{h}$.
}
\label{fig:TSseparaterh}
\end{figure}
From Fig.~\ref{fig:TSseparaterh}, it can be seen that for small values of the
event horizon radius $r_{h}$, we have two branches of solutions, one with a low
entropy and one with a much higher entropy.
Above about $r_{h}\sim 0.35$, we have just one branch of solutions.
\end{widetext}

\subsection{Stability of solutions}
\label{sec:stability}

We now investigate the key question of whether our solitons and black holes
with a non-minimally coupled scalar field are stable.
We consider spherically symmetric perturbations of the metric and scalar field.
The algebra rapidly becomes somewhat unwieldy, even with a computer algebra package,
so to simplify matters we follow the approach of \cite{raduew,ew2005} and use the
conformal transformation, described in Sec.~\ref{sec:conftrans},
to derive the perturbation equations with a minimally coupled scalar field, then
transforming back to the frame with a non-minimally coupled scalar field.

As usual, the metric perturbations can be eliminated to yield a single perturbation
equation for the quantity $\Psi $, which is defined in terms of the scalar field
perturbation $\delta \phi $ as follows:
\begin{equation}
\Psi =
r^{\frac {1}{2} \left( n-2 \right) }
\Omega ^{-\frac {1}{2}} {\mathcal {B}}^{\frac {1}{2}} \delta \phi ,
\label{eq:Psidef}
\end{equation}
where $\Omega =1-\xi \phi ^{2}$ as in Sec.~\ref{sec:conftrans} and
\begin{equation}
{\mathcal {B}} = \Omega +  \frac{4(n-1)}{n-2}\xi^2 \phi^2.
\label{eq:Bdef}
\end{equation}
The perturbation equation for time-periodic perturbations
$\Psi (t,r)=e^{i\sigma t}\Psi (r)$
takes the standard Schr\"odinger form
\begin{equation}
\sigma ^{2} \Psi = - \frac {d^{2}}{dr_{*}^{2}} \Psi + {\mathcal {U}} \Psi ,
\label{eq:schrodinger}
\end{equation}
where we have introduced the usual `tortoise' co-ordinate $r_{*}$ by
\begin{equation}
\frac {dr_{*}}{dr} = \frac {1}{He^{\delta }},
\label{eq:tortoise}
\end{equation}
and the perturbation potential ${\mathcal {U}}$ takes the lengthy form
\begin{widetext}
\begin{eqnarray}
{\mathcal {U}} & = & - \frac{H e^{2 \delta}}{r^2} \left\{
\frac{4 r^2}{(n-2)^2}\Omega^{-2} {\mathcal {A}}^2 H
- (n-2)(n-3)\frac{k}{2}
+ r^2 \Omega^{-1} \Lambda
- {\mathcal {B}}^{-1}r^2 \left[
\frac{2 \xi n \Lambda}{n-2} + \frac{8 \xi^2 n (n-1)}{(n-2)^2}
\phi^2 \Omega ^{-1} \Lambda \right]
\right. \nonumber \\ & &  \left.
  + \frac{2 {\mathcal {B}}^{-2} r^2 \xi n \phi \Lambda \Omega^{-1}}{n-2}
  \left[ \Omega \xi \phi + \frac{4(n-1)}{n-2} \xi^2 \phi(1 + \xi \phi^2) \right]
  + 2 \xi n \phi \Lambda \Omega^{-1} r^2 \phi' {\mathcal {A}}^{-1}
\right. \nonumber \\ & & \left.
  + \frac{(n-2)^3(n-3) k}{8} \phi'^2 {\mathcal {A}}^{-2}{\mathcal {B}}
- \frac{(n-2)^2}{4} r^2 \Omega^{-1} \Lambda \phi'^2 {\mathcal {A}}^{-2} {\mathcal {B}}
\right\} ,
\label{eq:PP}
\end{eqnarray}
\end{widetext}
with $\Omega $ and ${\mathcal {B}}$ defined above and
\begin{equation}
{\mathcal {A}} = \xi \phi \phi' - \frac{(n-2)}{2 r}\Omega .
\end{equation}
When $n=4$, the perturbation potential (\ref{eq:PP}) reduces to that in \cite{ew2005},
as expected.

\begin{figure}
\includegraphics[width=8cm]{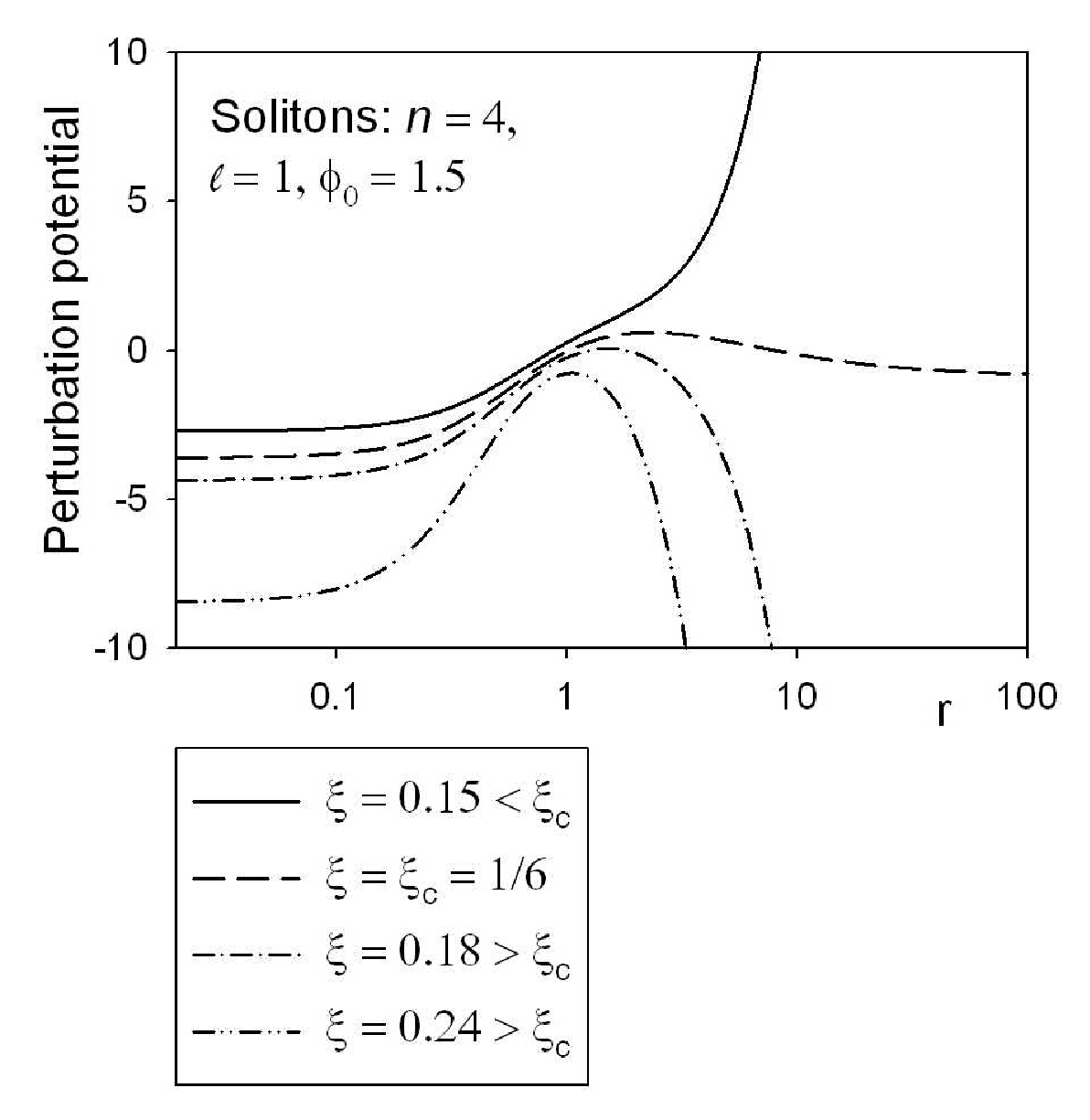}
\caption{
Perturbation potential ${\mathcal {U}}$ (\ref{eq:PP}) plotted as a function of
$r$ for some four-dimensional soliton solutions with $\ell = 1$, $\phi _{0}=1.5$
and four different values of the coupling constant $\xi $.
In this case the potential is finite at the origin.
The behaviour of the potential at infinity is in accordance with
(\ref{eq:PPasrtoinfinity}).
}
\label{fig:4dsolPP}
\end{figure}

\begin{figure}
\includegraphics[width=6cm,angle=270]{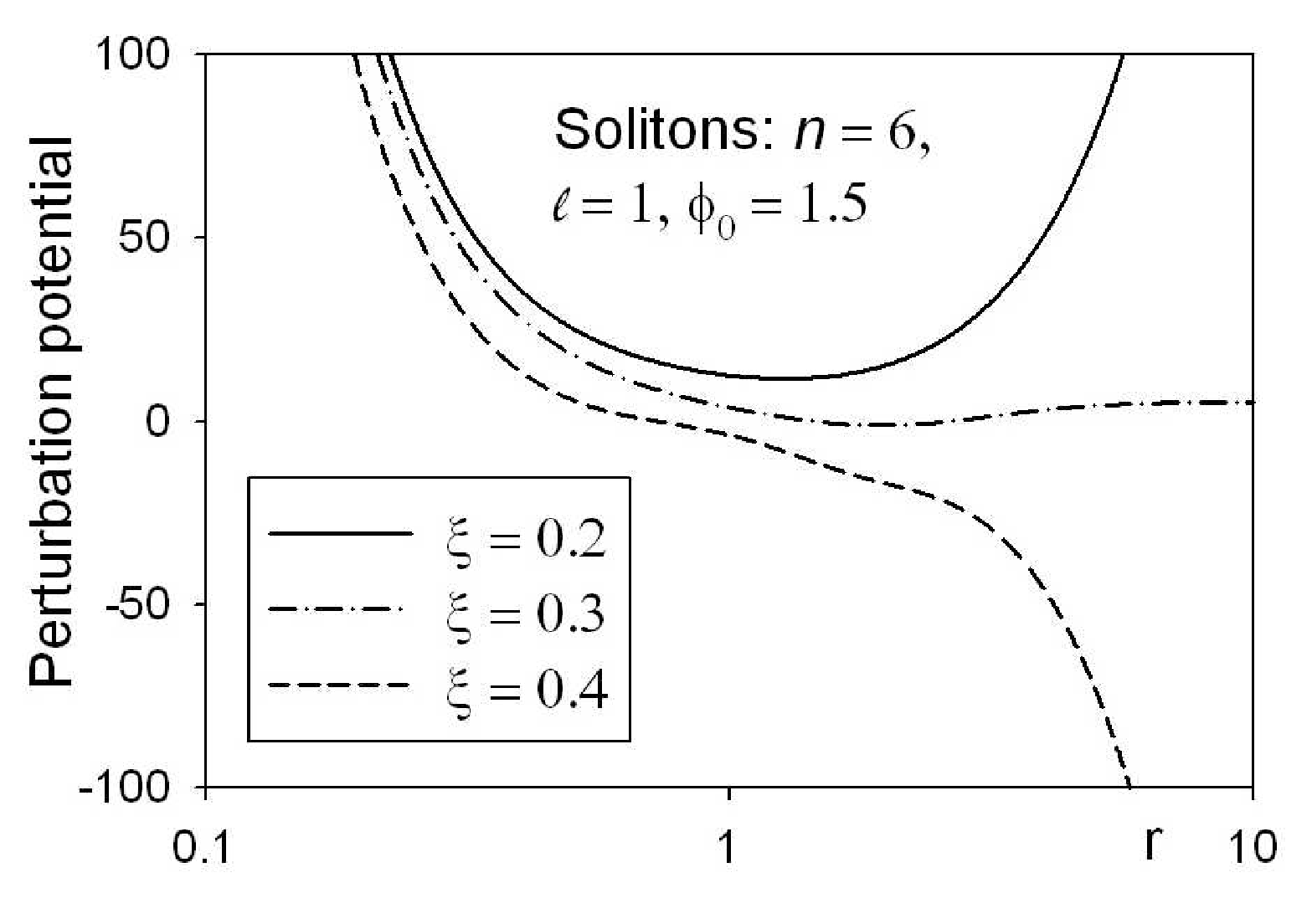}
\caption{
Perturbation potential ${\mathcal {U}}$ (\ref{eq:PP}) plotted as a function of
$r$ for some six-dimensional soliton solutions with $\ell = 1$, $\phi _{0}=1.5$
and three different values of the coupling constant $\xi $.
In this case the potential diverges at the origin, as expected from
equation (\ref{eq:PPasrto0}).
The behaviour of the potential at infinity is as predicted in
(\ref{eq:PPasrtoinfinity}).
}
\label{fig:6dsolPP}
\end{figure}

It is helpful to derive the behaviour of the perturbation potential at $r=0$,
$r=r_{h}$ and as $r\rightarrow \infty $.
Near the origin, using the boundary conditions (\ref{eq:rto0}), we have
\begin{equation}
{\mathcal {U}} =  \frac{e^{2 \delta_0}}{r^2}
\left[  \frac{(n-2)(n-3)}{2} - 1 \right]
+ O(1).
\label{eq:PPasrto0}
\end{equation}
This means that the perturbation potential ${\mathcal {U}}$ diverges to $+\infty $
at the origin
unless $n=4$, as can be seen by comparing the potentials in Figs.~\ref{fig:4dsolPP}
and \ref{fig:6dsolPP}.

\begin{figure}
\includegraphics[width=6cm,angle=270]{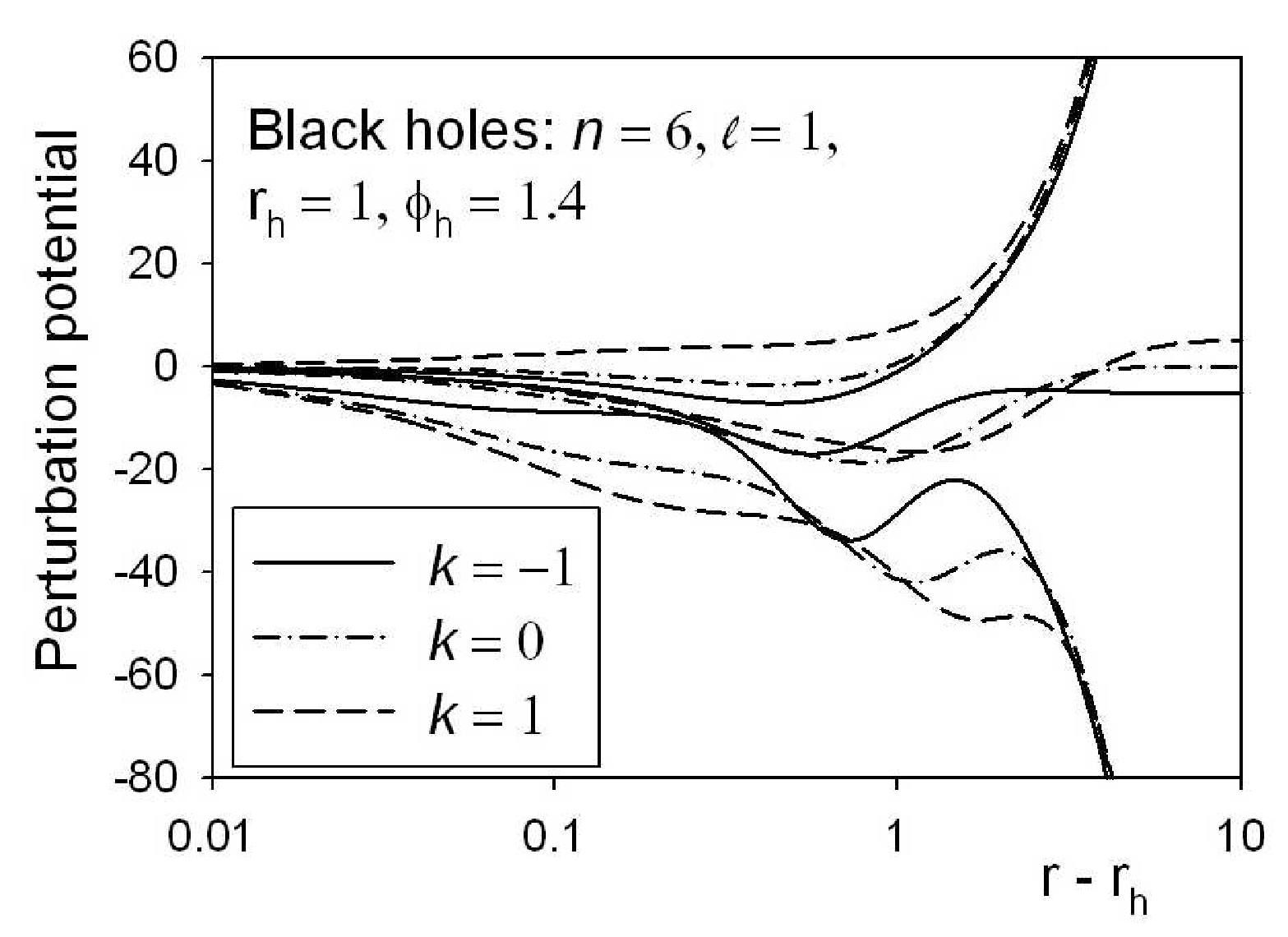}
\caption{
Perturbation potential ${\mathcal {U}}$ (\ref{eq:PP}) plotted as a function of
$r$ for some six-dimensional black hole solutions with $\ell = 1$,
$r_{h}=1$ and $\phi _{h}=1.4$.
For each value of $k$, we plot the potential for $\xi = 0.2$, $0.3$ and $0.4$.
The potential vanishes at the event horizon, and the behaviour at infinity
is governed by (\ref{eq:PPasrtoinfinity}).
The potentials for $\xi = 0.2$ diverge to $\infty $ as $r\rightarrow \infty $,
those for $\xi =0.3$ converge to a finite limit as $r\rightarrow \infty $,
while those for $\xi = 0.4$ diverge to $-\infty $ as $r\rightarrow \infty $.
}
\label{fig:6dbhPP}
\end{figure}

At the event horizon, $H(r_{h})=0$ and all other quantities are finite, so it is
straightforward to see that the potential ${\mathcal {U}}$ vanishes there,
as illustrated in Fig.~\ref{fig:6dbhPP}.

As $r\rightarrow \infty $, using the boundary conditions
(\ref{eq:phiinf}, \ref{eq:rinfdelta}, \ref{eq:rinfH}), we find
\begin{eqnarray}
{\mathcal {U}} & = &
\frac {2n\Lambda ^{2} r^{2}}{(n-2)^{2}(n-1)^{2}} \left[
 \left( n - 3 \right)
- 2\xi \left( n-1 \right)
\right]
\nonumber \\ & &
+ \frac {2k\Lambda }{(n-2)(n-1)}\left[
\xi n (n-1) - n^{2} + 4n - 2
\right]
\nonumber \\ & &
+ O\left( r^{-2} \right) .
\label{eq:PPasrtoinfinity}
\end{eqnarray}
Therefore the potential ${\mathcal {U}}$ diverges as $r\rightarrow \infty $ unless
$\xi = \left( n-3 \right) / [2(n-1)]$, which corresponds to $\xi = \xi _{c}=1/6$ when
$n=4$.
This is confirmed by the example potentials plotted in
Figs.~\ref{fig:4dsolPP}--\ref{fig:6dbhPP}.

As in \cite{raduew}, we find that in general the perturbation potential
has a complex dependence on $n$, $\phi _{0}$ or $\phi _{h}$ as applicable, $r_{h}$,
$k$ and $\xi $  (see particularly Fig.~\ref{fig:6dbhPP}).
In some cases, the perturbation
potential ${\mathcal {U}}$ is positive everywhere and we can immediately deduce that
the solutions are stable.
From (\ref{eq:PPasrtoinfinity}), this is only possible for
$\xi \le \left( n-3 \right) / [2(n-1)]$.

\begin{figure}
\includegraphics[width=6cm,angle=270]{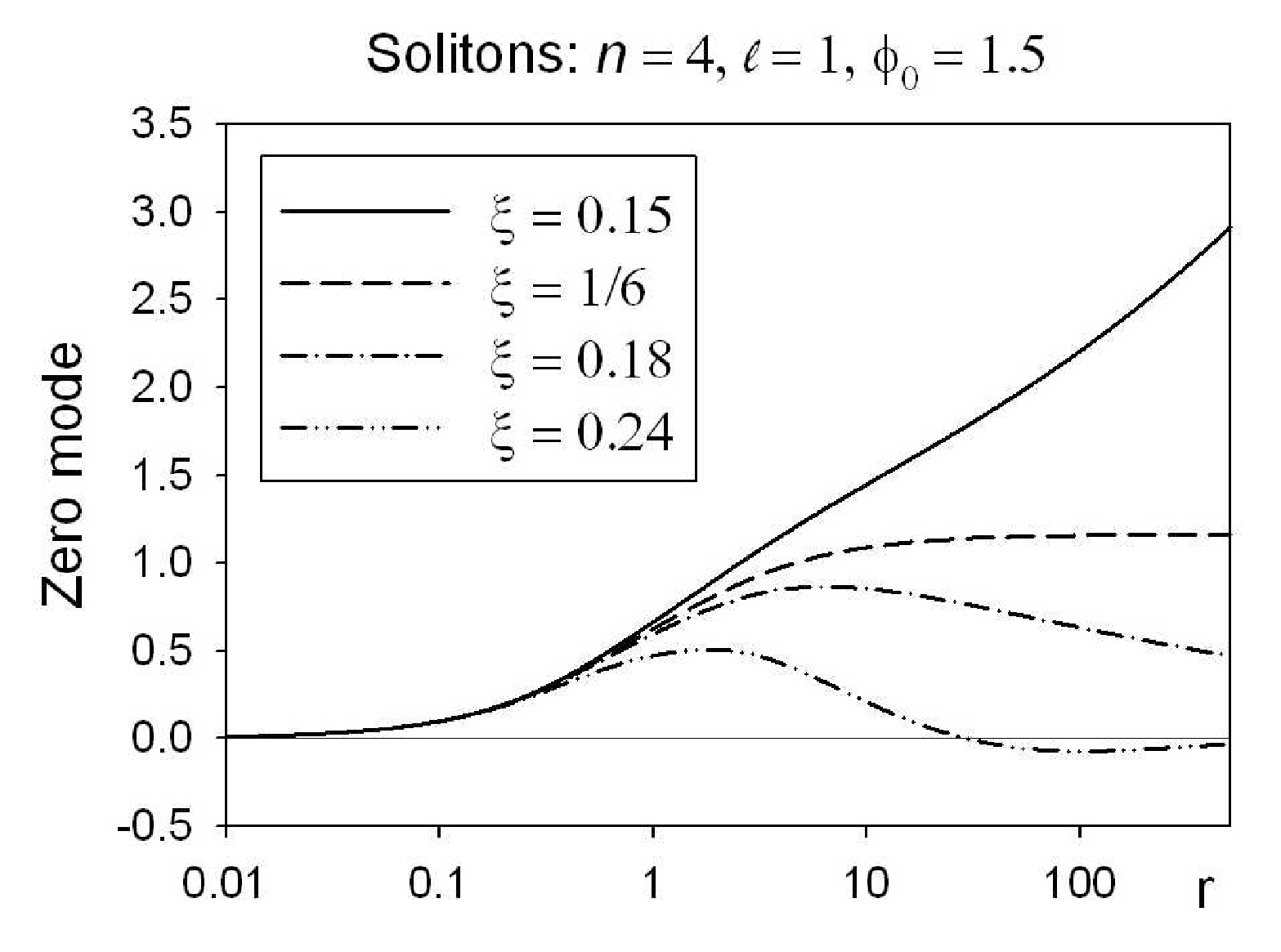}
\caption{
Zero mode $\Psi _{0}$ plotted as a function of
$r$ for some four-dimensional soliton solutions with $\ell = 1$, $\phi _{0}=1.5$
and four different values of the coupling constant $\xi $.
}
\label{fig:4dsolZM}
\end{figure}

\begin{figure}
\includegraphics[width=6cm,angle=270]{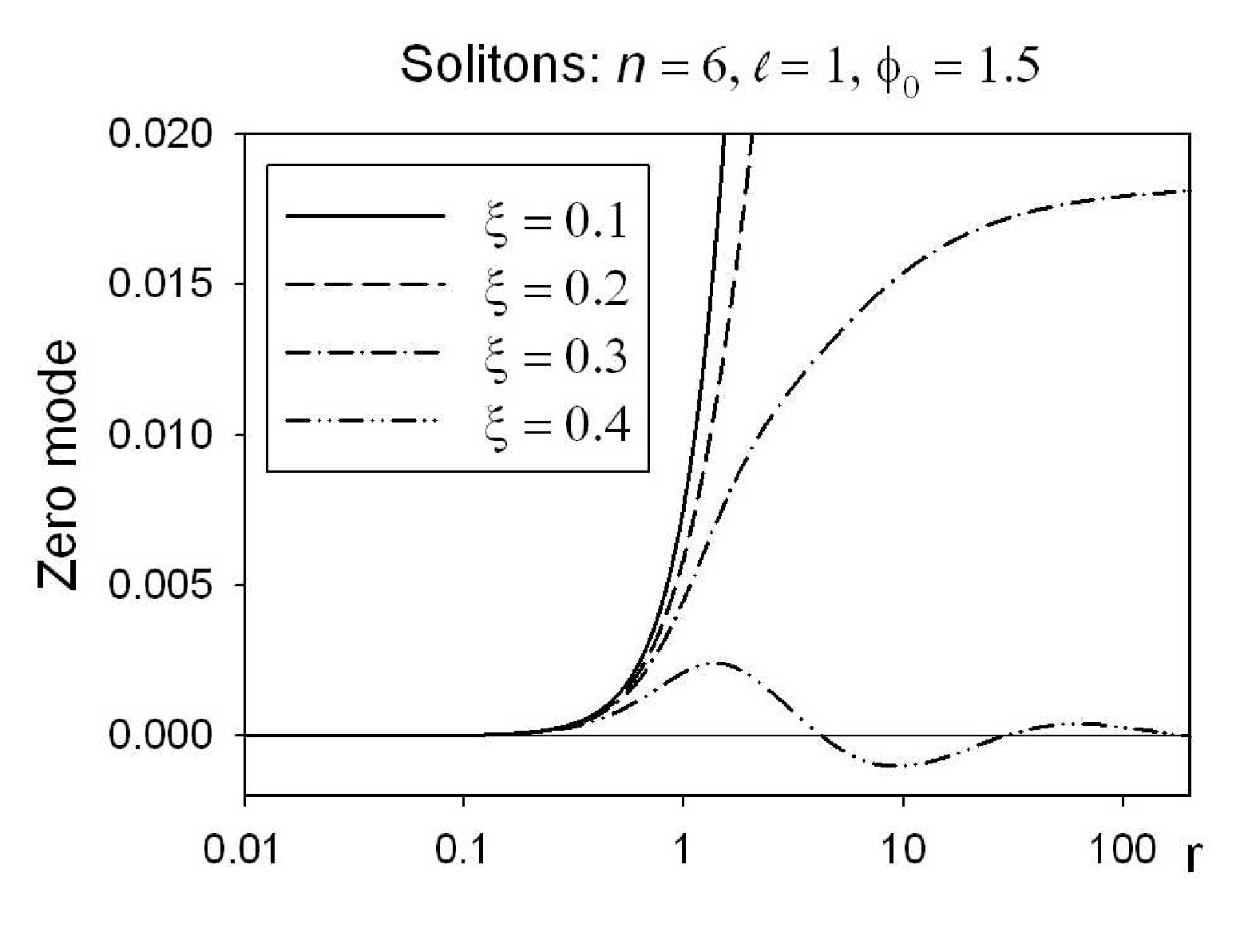}
\caption{
Zero mode $\Psi _{0}$ plotted as a function of
$r$ for some six-dimensional soliton solutions with $\ell = 1$, $\phi _{0}=1.5$
and four different values of the coupling constant $\xi $.
}
\label{fig:6dsolZM}
\end{figure}

\begin{figure}
\includegraphics[width=6cm,angle=270]{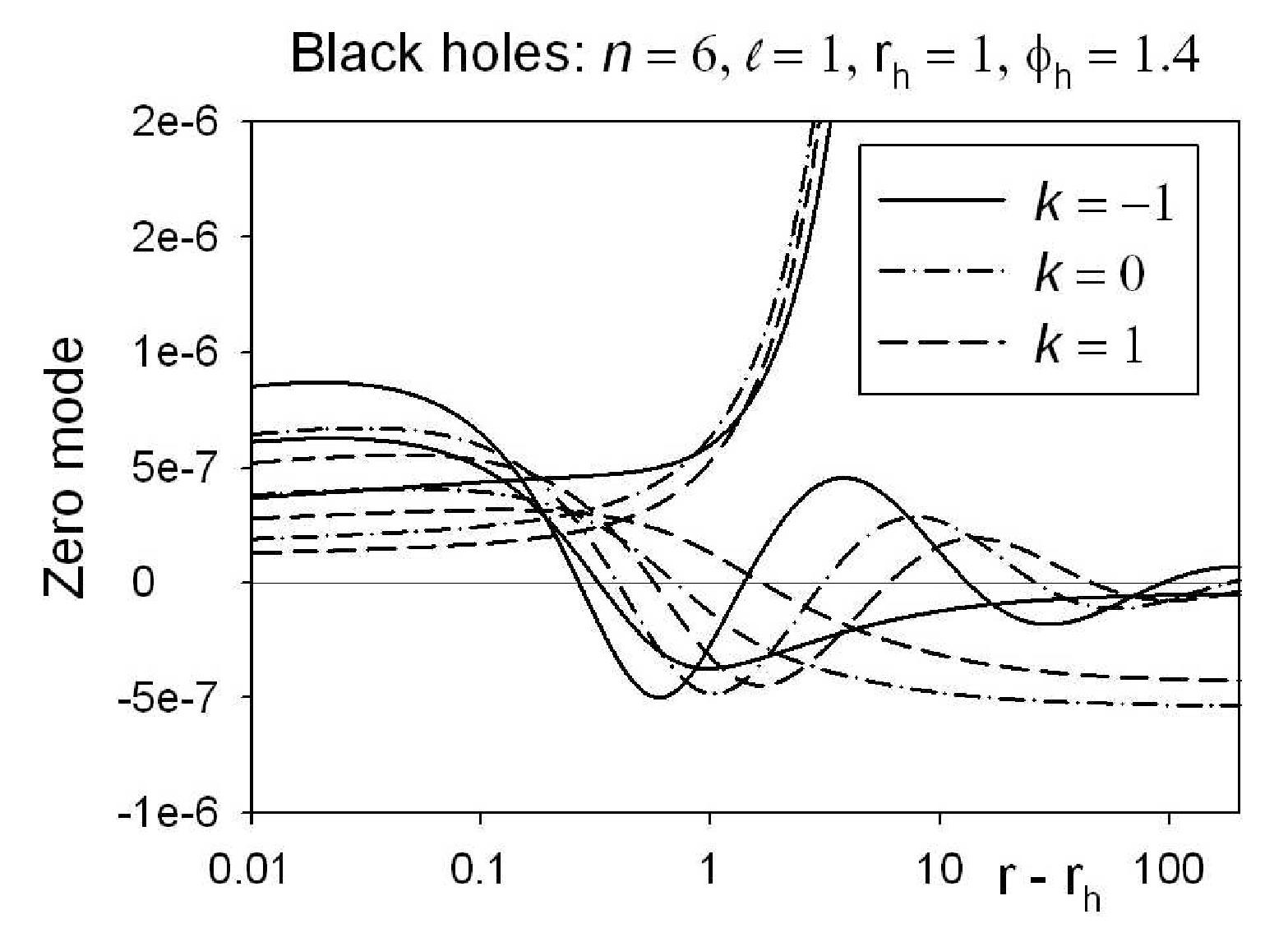}
\caption{
Zero mode $\Psi _{0}$ plotted as a function of
$r$ for some six-dimensional black hole solutions with $\ell = 1$,
$r_{h}=1$ and $\phi _{h}=1.4$.
For each value of $k$, we plot the zero mode for $\xi = 0.1$, $0.3$ and $0.4$.
The zero modes for $\xi =0.1$ diverge to $+\infty $ as $r\rightarrow \infty $;
those for $\xi = 0.3$ have a single zero; and those for $\xi = 0.4$ oscillate
about $0$ as $r\rightarrow \infty $.
}
\label{fig:6dbhZM}
\end{figure}

To study the stability of the solutions for which the perturbation potential
${\mathcal {U}}$ is not everywhere positive, we follow the approach of
\cite{raduew,ew2005}, and consider the zero mode perturbation $\Psi _{0}$, that is,
the solution of the perturbation equation (\ref{eq:schrodinger}) with $\sigma = 0$.
Near $r=0$, from the differential equation (\ref{eq:schrodinger}), we find that
$\Psi _{0}= O(r^{\alpha })$, where
\begin{equation}
\alpha = \frac{1}{2} \pm \frac{1}{2} {\sqrt{2 n^2 - 10 n + 9}},
\label{eqn:alphavalue}
\end{equation}
and we chose the positive sign so that $\Psi _{0}$ is regular at the origin.
Near $r=r_{h}$, we simply require $\Psi =O(r-r_{h})$, and as $r \rightarrow \infty $,
we have $\Psi _{0}= O(r^{-\beta })$, where
\begin{equation}
\beta = \frac {1}{2} \pm \frac {1}{2}
{\sqrt {1 +2n(n-3) - 4n(n-1) \xi }} .
\label{eq:betavalue}
\end{equation}
The behaviour of $\Psi _{0}$ as $r\rightarrow \infty $  is therefore complicated.
For $\xi < [1+ 2n(n-3)]/[4n(n-1)]$, the constant $\beta $ is real, and the dominant
behaviour of $\Psi _{0}$ will be from taking the negative sign in (\ref{eq:betavalue}).
This value of $\beta $ will be positive (corresponding to $\Psi _{0}\rightarrow 0 $
as $r\rightarrow \infty $) as long as $\xi > (n-3)/[2(n-1)]$, but is negative
(corresponding to $\Psi _{0}$ diverging as $r\rightarrow \infty $) for
$\xi < (n-3)/[2(n-1)]$.
For $\xi > [1+2n(n-3)]/[4n(n-1)]$, the constant $\beta $ is complex, with positive real
part.
In this case $\Psi _{0}\rightarrow 0$ as $r\rightarrow \infty $,
but $\Psi _{0}$ is oscillating as $r\rightarrow \infty $.
Integrating the perturbation equation (\ref{eq:schrodinger}) over the original radial
co-ordinate $r$, from $r=0$ or $r=r_{h}$, as applicable, gives examples of the typical
behaviour of the zero modes, which are shown in
Figs.~\ref{fig:4dsolZM}--\ref{fig:6dbhZM}.

For $0< \xi < \xi _{c}$, we find that the zero mode $\Psi _{0}$ has no zeros,
and from this we can deduce that there can be no negative eigenvalues $\sigma ^{2}<0$
of the perturbation equation (\ref{eq:schrodinger}), in other words, the solutions
are stable.
For $\xi > [1+2n(n-3)]/[4n(n-1)]$, we find, as expected, that the zero mode $\Psi _{0}$
oscillates many times as $r\rightarrow \infty $, leading us to conclude that there
is at least one negative eigenvalue $\sigma ^{2}<0$ of the perturbation equation
(\ref{eq:schrodinger}), and the solutions are unstable.
For $\xi _{c}<\xi <[1+2n(n-3)]/[4n(n-1)]$, the situation is more complicated.
We find for some solutions that the zero mode $\Psi _{0}$ has no zeros,
indicating the stability of the solutions; but
for others it has at least one zero and we conclude that the solutions are
unstable.

\begin{figure}
\includegraphics[width=7.5cm]{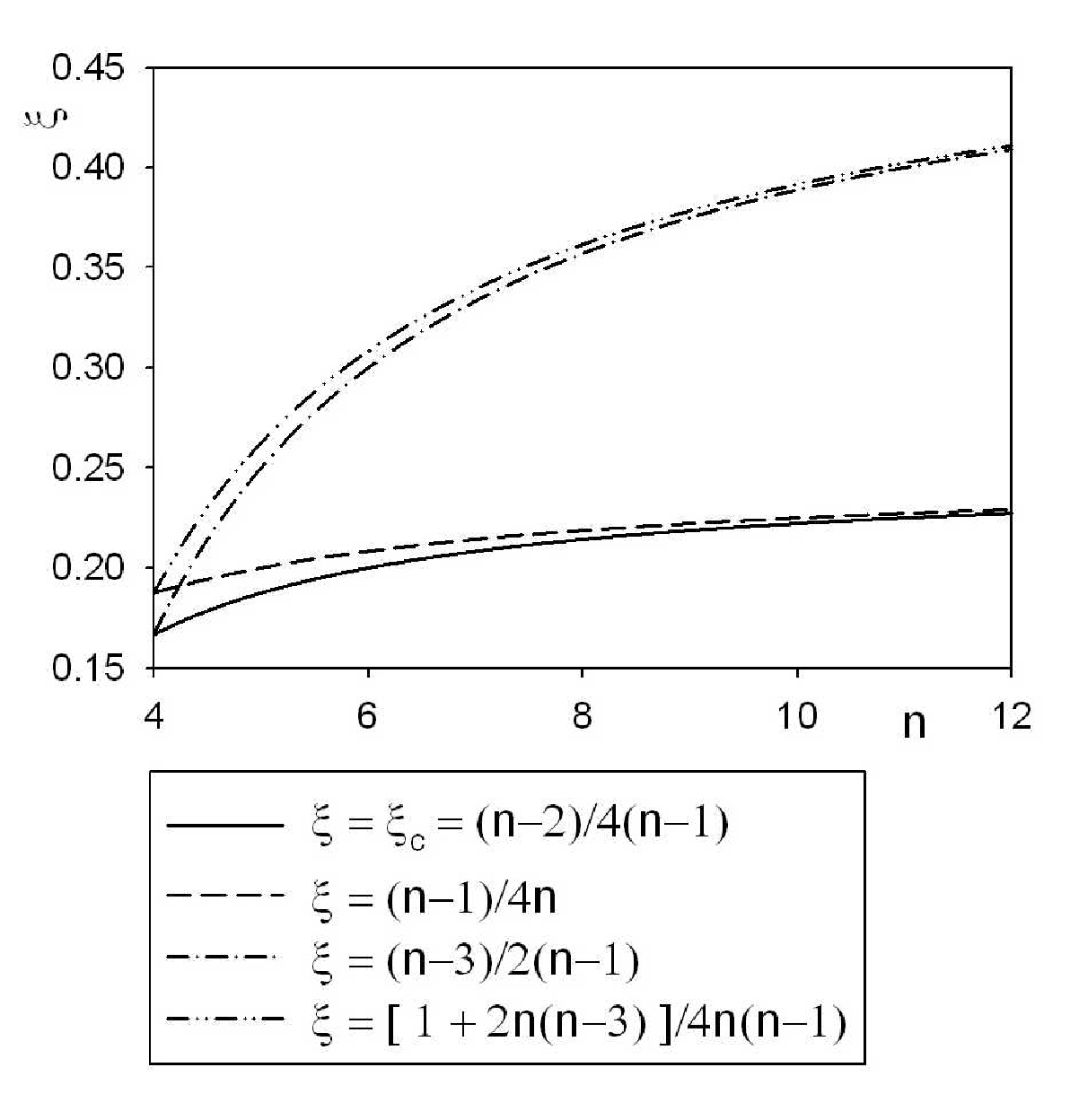}
\caption{The key values of the coupling constant $\xi $
discussed in the text, as
functions of the number of dimensions $n$.}
\label{fig:xi}
\end{figure}

As discussed in Sec.~\ref{sec:boundary}, the
equilibrium scalar field oscillates as $r\rightarrow \infty $
if $\xi > (n-1)/(4n)$ (\ref{eq:p}).
Interestingly, this coincides with the value of $\xi $ coming from the
Breitenlohner-Freedman bound in $n$ dimensions \cite{BFbound}, which
states that scalar fields in pure anti-de Sitter space are stable if their
mass satisfies the inequality
\begin{equation}
m_{BF}^{2} > \frac {\Lambda (n-1)}{2(n-2)}.
\label{eq:BFbound}
\end{equation}
In our case the ``effective'' mass of the scalar field is given by $\xi $ multiplied
by the value of the Ricci scalar curvature at infinity, which is $2\Lambda n/(n-2)$.
Bearing in mind that $\Lambda $ is negative, the inequality
(\ref{eq:BFbound}) then becomes $\xi < (n-1)/(4n)$.
For any $n>4$, this value of $\xi $ is always greater than $\xi _{c}$, as
can be seen in Fig.~\ref{fig:xi}, but the gap between $\xi _{c}$ and the value
of $\xi $ coming from the Breitenlohner-Freedman bound narrows as $n$ gets large.
From the Breitenlohner-Freedman bound, we therefore might
expect that any non-trivial solutions with $\xi > (n-1)/(4n)$ are unstable.
We find that for $\xi > (n-3)/[2(n-1)]$, the perturbation
potential ${\cal {U}}$ diverges to $-\infty $ as $r\rightarrow \infty $.
Analysis of the zero mode $\Psi _{0}$ reveals that it
oscillates as $r\rightarrow \infty $ only if
$\xi > \left[ 1 + 2n(n-3) \right] /[4n(n-1)]$.
These latter two values of $\xi $ are also shown in Fig.~\ref{fig:xi}.
The value of $\xi $ at which the potential diverges to $-\infty $ at infinity
is equal to $\xi _{c}$ only when $n=4$, and for $n>4$ it is considerably larger
than both $\xi _{c}$ and the value of $\xi $ coming from the Breitenlohner-Freedman
bound.
The value of $\xi $ at which the zero mode oscillates equals the value of $\xi $
coming from the Breitenlohner-Freedman bound when $n=4$, but for all $n\ge 4$
it is greater than the value of $\xi $ at which the perturbation potential diverges
to $-\infty $.

For $n>4$, we therefore have a complicated picture.
For all the solutions we have studied, we find that they are stable when
$0<\xi <\xi _{c}$, in accordance with the results in \cite{ew2005} for $n=4$.
All the solutions we have studied are unstable for
$\xi > \left[ 1 + 2n(n-3) \right] /[4n(n-1)]$.
However, while the Breitenlohner-Freedman bound would lead us to expect that
all the solutions with $\xi > (n-1)/(4n)$ would be unstable, in practice
we are unable to reach any general conclusions about the stability
of solutions when $(n-1)/(4n) < \xi < \left[ 1 + 2n(n-3) \right] /[4n(n-1)]$,
with some solutions being stable and some unstable.

\section{Conclusions}
\label{sec:conc}

In this paper we have studied the existence of soliton and black hole solutions
of the Einstein equations in the presence of a cosmological constant and
a non-minimally coupled scalar field with zero self-interaction potential.
We have considered all space-time dimensions greater than or equal to four, and
topological black holes as well as the usual spherically symmetric solitons and black
holes, covering all the possibilities not previously considered in the literature
\cite{raduew,ew2005}.

Since we have a zero self-interaction potential, we are able to use elementary
arguments to show the non-existence of soliton or black hole solutions, except
when the cosmological constant is negative and the coupling constant $\xi $ (which
governs the coupling between the scalar field and the Ricci scalar curvature)
is positive, when
we find non-trivial soliton and hairy black hole solutions.
The field equations are highly complex and so we have only considered numerical
solutions in this paper.
It is likely that solution generating techniques \cite{higherdimgen} could be used
to give analytic solutions, but only with non-zero self-interaction potentials.

The behaviour of the solutions as $\xi $ varies is rather more complex in
more than four dimensions than it was in four space-time dimensions \cite{ew2005}.
If $\xi >0$ but less than the value for conformal coupling,
for all the numerical solutions we studied, the scalar field was monotonically
decreasing to zero from its value at the origin or
the black hole event horizon, as applicable, and, furthermore, all
such solutions that we studied were stable.
We should emphasize that we have only considered spherically symmetric perturbations
in this paper, and that while these solutions are stable under such perturbations, it is possible that there exist non-spherically symmetric unstable modes.

For values of $\xi $ greater than that for conformal coupling, the picture
is less straightforward.
The scalar field is oscillating for $\xi > (n-1)/(4n)$, in accordance with the
Breitenlohner-Freedman \cite{BFbound} bound for scalar fields in pure anti-de Sitter
space.
This would lead us to expect that all solutions for $\xi > (n-1)/(4n)$ would be unstable.
However,  while all the solutions we have studied with
$\xi > \left[ 1 + 2n(n-3) \right] /[4n(n-1)]$ have been unstable, we have found
conflicting results for $(n-1)/(4n) < \xi < \left[ 1 + 2n(n-3) \right] /[4n(n-1)]$,
with some solutions being stable and some unstable.

The oscillatory nature of the scalar field for $\xi > (n-1)/(4n)$ also leads to
oscillations in the metric function $m(r)$, whose limit as $r\rightarrow \infty $
one would normally take as a measure of the mass of the solution.
We leave the question of how to define the mass when $m(r)$ is oscillating for
future work.
In the absence of an appropriate definition of mass, we have been unable to
perform a complete thermodynamic analysis of our solutions, which would perhaps help to
resolve the stability issue outlined above.
We hope to return to these questions.

\begin{widetext}
Our results are summarized in the updated Tab.~\ref{tab:KnowledgeForV0v3},
where the new results in this paper are highlighted in bold.
\begin{table}[h]
\begin{tabular}[c]{|c||c|c|c|c|c|}
\hline
$n>4$ & $\xi < 0$ & $\xi = 0$ & $0 < \xi < \xi_c$ & $\xi = \xi_c$ & $\xi > \xi_c$ \\
\hline
\hline
$\Lambda > 0$
& {\bf {no solutions}}
& {\bf {no solutions}}
& {\bf {no solutions}}
& {\bf {no solutions}}
& {\bf {no solutions}}
\\
\hline
$\Lambda = 0$
& {\bf {no solutions}}
& {\bf {no solutions}}
& no solutions \cite{saa}
& {\bf {no solutions}}
& {\bf {no solutions}}
\\
\hline
$\Lambda < 0$
& {\bf {no solutions}}
& {\bf {no solutions}}
& {\bf {stable solutions}}
& stable solutions \cite{raduew}
& {\bf {mostly unstable solutions}}
\\
\hline
\end{tabular}
\caption{Summary of existence and non-existence of soliton and black hole
solutions in more than four space-time dimensions.  }
\label{tab:KnowledgeForV0v3}
\end{table}
\end{widetext}

\begin{acknowledgments}
The work of DH is supported by the Nuffield Foundation (UK), reference number
URB/35440.
The work of EW is supported by STFC (UK), grant number ST/G000611/1.
\end{acknowledgments}


\begin{thebibliography}{99}

\bibitem{heuslerreview}
M.~Heusler, {\it {Black Hole Uniqueness Theorems}} (Cambridge University Press,
Cambridge, 1996).

\bibitem{emparanreall}
R.~Emparan and H.~S.~Reall,  Living Rev.\ Rel.\  {\bf {11}}, 6 (2008).

\bibitem{bekreview}
J.~D.~Bekenstein, {\eprint {arXiv:gr-qc/9605059}}.

\bibitem{mincoupleduniqueness}
J.~D.~Bekenstein, Phys.\ Rev.\  D {\bf {5}}, 1239 (1972);
 D {\bf {5}}, 2403 (1972);
 D {\bf {51}}, R6608 (1996);
\newline
T.~Hertog,  Phys.\ Rev.\  D {\bf {74}}, 084008 (2006);
\newline
M.~Heusler, Class.\ Quantum Gravity {\bf {12}}, 779 (1995);
\newline
M.~Heusler and N.~Straumann, Class.\ Quantum Gravity {\bf {9}}, 2177 (1992);
\newline
D.~Sudarsky, Class.\ Quantum Gravity {\bf {12}}, 579 (1995).

\bibitem{mincoupnum}
U.~Nucamendi and M.~Salgado,  Phys.\ Rev.\  D {\bf {68}}, 044026 (2003);
\newline
M.~Alcubierre, J.~A.~Gonzalez and M.~Salgado, Phys.\ Rev.\  D {\bf {70}}, 064016 (2004);
\newline
A.~Corichi, U.~Nucamendi and M.~Salgado,  Phys.\ Rev.\  D {\bf {73}}, 084002 (2006).

\bibitem{mincoupan}
O.~Bechmann and O.~Lechtenfeld,  Class.\ Quant.\ Grav.\  {\bf {12}}, 1473 (1995);
\newline
K.~A.~Bronnikov and G.~N.~Shikin,  Grav.\ Cosmol.\  {\bf {8}}, 107 (2002);
\newline
H.~Dennhardt and O.~Lechtenfeld,  Int.\ J.\ Mod.\ Phys.\  A {\bf {13}}, 741 (1998).

\bibitem{vacgen}
H.~A.~Buchdahl, Phys.\ Rev.\ {\bf {115}}, 1325 (1959);
\newline
A.~I.~Janis, D.~C.~Robinson and J.~Winicour,  Phys.\ Rev.\  {\bf {186}}, 1729 (1969);
\newline
D.~L.~Wiltshire,  Phys.\ Rev.\  D {\bf {46}}, 5682 (1992).

\bibitem{toriidS}
T.~Torii, K.~Maeda and N.~Narita, Phys.\ Rev.\ D {\bf {59}}, 064027 (1999).

\bibitem{toriiadS}
D.~H.~Park,  Class.\ Quant.\ Grav.\  {\bf {25}}, 095002 (2008);
\newline
D.~Sudarsky and J.~A.~Gonzalez,  Phys.\ Rev.\  D {\bf {67}}, 024038 (2003);
\newline
T.~Torii, K.~Maeda and N.~Narita, Phys.\ Rev.\ D {\bf {64}}, 044007 (2001).

\bibitem{zlosh}
K.~G.~Zloshchastiev,  Phys.\ Rev.\ Lett.\  {\bf {94}}, 121101 (2005).

\bibitem{lahiri}
S.~Bhattacharya and A.~Lahiri,  Phys.\ Rev.\ Lett.\  {\bf {99}}, 201101 (2007).

\bibitem{mtzminimal}
C.~Martinez, R.~Troncoso and J.~Zanelli,  Phys.\ Rev.\  D {\bf {70}}, 084035 (2004).

\bibitem{farakos}
K.~Farakos, A.~P.~Kouretsis and P.~Pasipoularides, Phys.\ Rev.\ D {\bf {80}}, 064020
(2009).

\bibitem{zeng}
D.-f.~Zeng, \eprint{arXiv:0903.2620 [hep-th]}.

\bibitem{adSCFThair}
A.~Buchel and C.~Pagnutti, Nucl.\ Phys.\ {\bf {B824}}, 85 (2010);
\newline
S.~S.~Gubser, A.~Nellore, S.~S.~Pufu and F.~D.~Rocha,
  Phys.\ Rev.\ Lett.\  {\bf {101}}, 131601 (2008).

\bibitem{SUGRAhair}
T.~Hertog and K.~Maeda,  JHEP {\bf {0407}}, 051 (2004).

\bibitem{SUGRAhairstab}
T.~Hertog and K.~Maeda,  Phys.\ Rev.\  D {\bf 71}, 024001 (2005).

\bibitem{BBMB}
J.~D.~Bekenstein, Ann.\ Phys.\ (NY) {\bf {82}}, 535 (1974); {\bf {91}}, 75 (1975);
\newline
N.~M.~Bocharova, K.~A.~Bronnikov and V.~N.~Mel'nikov, Vestn.\ Mosk.\ Univ.\ Fiz.\
{\bf {25}}, 706 (1970).

\bibitem{BBMBunique}
B.~C.~Xanthopoulos and T.~Zannias, J.\ Math.\ Phys.\ {\bf {32}}, 1875 (1991).

\bibitem{BBMBunstable}
K.~A.~Bronnikov and Y.~N.~Kireyev, Phys.\ Lett.\ A {\bf {67}}, 95 (1978).

\bibitem{conformalnonzeroV}
P.~I.~Kuriakose and V.~C.~Kuriakose, \eprint{arXiv:0805.4554 [gr-qc]}.

\bibitem{mtz}
C.~Martinez, R.~Troncoso and J.~Zanelli, Phys.\ Rev.\ D {\bf {67}}, 024008 (2003).

\bibitem{harper}
G.~Dotti, R.~J.~Gleiser and C.~Martinez,  Phys.\ Rev.\  D {\bf {77}}, 104035 (2008);
\newline
T.~J.~T.~Harper, P.~A.~Thomas, E.~Winstanley and P.~M.~Young, Phys.\ Rev.\ D
{\bf {70}}, 064023 (2004).

\bibitem{raduew}
E.~Radu and E.~Winstanley, Phys.\ Rev.\ D {\bf {72}}, 024017 (2005).

\bibitem{ew2003}
E.~Winstanley, Found.\ Phys.\ {\bf {33}}, 111 (2003).

\bibitem{nonminnohair}
E.~Ayon-Beato,  Class.\ Quantum Gravity  {\bf {19}}, 5465 (2002);
\newline
A.~E.~Mayo and J.~D.~Bekenstein, Phys.\ Rev.\ D {\bf {54}}, 5059 (1996);
\newline
A.~Saa, J.\ Math.\ Phys.\ {\bf {37}}, 2346 (1996).

\bibitem{saa}
A.~Saa,
Phys.\ Rev.\ D {\bf {53}}, 7377 (1996).

\bibitem{ew2005}
E.~Winstanley, Class.\ Quantum Gravity {\bf {22}}, 2233 (2005).

\bibitem{pena}
I.~Pena and D.~Sudarsky, Class.\ Quantum Gravity {\bf {18}}, 1461 (2001).

\bibitem{BFbound}
P.~Breitenlohner and D.~Z.~Freedman, Phys.\ Lett.\ B {\bf {115}}, 197 (1982);
Ann.\ Phys.\ (NY) {\bf {144}}, 249 (1982).

\bibitem{topological}
D.~Birmingham, Class.\ Quantum Gravity {\bf {16}}, 1197 (1999);
\newline
J.~P.~S.~Lemos, 
Class.\ Quantum Gravity {\bf {12}}, 1081 (1995);
Phys.\ Lett.\ B {\bf {353}}, 46 (1995);
\newline
J.~P.~S. Lemos and V.~T. Zanchin, Phys.\ Rev.\ D {\bf {54}}, 3840 (1996);
\newline
L.~Vanzo, Phys.\ Rev.\ D {\bf {56}}, 6475 (1997).

\bibitem{4dtopnegcurvmincoup}
C.~Martinez, R.~Troncoso and J.~Zanelli,  Phys.\ Rev.\  D {\bf {70}}, 084035 (2004).

\bibitem{higherdimgen}
K.~Tangen, \eprint{arXiv:0705.4372 [gr-qc]};
\newline
I.~K.~Wehus and F.~Ravndal,  J.\ Phys.\ Conf.\ Ser.\  {\bf {66}}, 012024 (2007);
\newline
B.~C.~Xanthopoulos and T.~Zannias, Phys.\ Rev.\ D {\bf {40}}, 2564 (1989).

\bibitem{klimcik}
C.~Klimcik, J.\ Math.\ Phys.\ {\bf {34}}, 1914 (1993);
\newline
B.~C.~Xanthopoulos and T.~E.~Dialynas, J.\ Math.\ Phys.\ {\bf {33}}, 1463 (1992).

\bibitem{3Dconf}
C.~Martinez and J.~Zanelli,  Phys.\ Rev.\  D {\bf {54}}, 3830 (1996).

\bibitem{nadalini}
M.~Nadalini, L.~Vanzo and S.~Zerbini,  Phys.\ Rev.\  D {\bf {77}}, 024047 (2008).

\bibitem{henneaux}
M.~Henneaux, C.~Martinez, R.~Troncoso and J.~Zanelli,
Phys.\ Rev.\  D {\bf {70}}, 044034 (2004);
Ann.\ Phys.\  {\bf {322}}, 824 (2007).

\bibitem{maeda}
K.-i.~Maeda, Phys.\ Rev.\ D {\bf {39}}, 3159 (1989).

\bibitem{otherconf}
A.~J.~Accioly, U.~F.~Wichoski, S.~F.~Kwok and N.~L.~P.~Pereira da Silva,
Class.\ Quantum Gravity {\bf {10}}, L215 (1993);
\newline
N.~Banerjee, S.~Sen and N.~Dadhich, Mod.\ Phys.\ Lett.\  A {\bf {16}}, 1223 (2001).

\bibitem{thermo}
A.~M.~Barlow, D.~Doherty and E.~Winstanley,  Phys.\ Rev.\  D {\bf {72}}, 024008 (2005);
\newline
E.~Winstanley, \eprint{arXiv:gr-qc/0408046};
\newline
O.~B.~Zavlaskii, Class.\ Quantum Gravity {\bf {19}}, 3783 (2002).

\bibitem{entropy}
A.~Ashtekar, A.~Corichi, and D.~Sudarsky, Class.\ Quantum Gravity {\bf {20}}, 3413
(2003);
\newline
V.~Iyer and R.~M.~Wald, Phys.\ Rev.\ D {\bf {50}}, 846 (1994).

%\bibitem{hp}
%S.~W.~Hawking and D.~N.~Page, Commun.\ Math.\ Phys.\ {\bf {87}}, 577  (1983).

\end{thebibliography}
\end{document}